\documentclass[aps,prl,twocolumn,showpacs,superscriptaddress]{revtex4}
\usepackage{graphicx}  
\usepackage{dcolumn}  
\usepackage{bm}    
\usepackage{amssymb}
\usepackage{epsfig}
\usepackage{subfigure}

\newcommand{\mgg}{\mbox{\ensuremath{M_{\gamma\gamma}}}}
\newcommand{\dphigg}{\mbox{\ensuremath{\Delta\phi_{\gamma\gamma}}}}

\begin{document}

\hspace{5.2in} \mbox{FERMILAB-PUB-11-331-E}

\title{Search for the standard model and a fermiophobic Higgs boson in diphoton final states} 
\affiliation{Universidad de Buenos Aires, Buenos Aires, Argentina}
\affiliation{LAFEX, Centro Brasileiro de Pesquisas F{\'\i}sicas, Rio de Janeiro, Brazil}
\affiliation{Universidade do Estado do Rio de Janeiro, Rio de Janeiro, Brazil}
\affiliation{Universidade Federal do ABC, Santo Andr\'e, Brazil}
\affiliation{Instituto de F\'{\i}sica Te\'orica, Universidade Estadual Paulista, S\~ao Paulo, Brazil}
\affiliation{Simon Fraser University, Vancouver, British Columbia, and York University, Toronto, Ontario, Canada}
\affiliation{University of Science and Technology of China, Hefei, People's Republic of China}
\affiliation{Universidad de los Andes, Bogot\'{a}, Colombia}
\affiliation{Charles University, Faculty of Mathematics and Physics, Center for Particle Physics, Prague, Czech Republic}
\affiliation{Czech Technical University in Prague, Prague, Czech Republic}
\affiliation{Center for Particle Physics, Institute of Physics, Academy of Sciences of the Czech Republic, Prague, Czech Republic}
\affiliation{Universidad San Francisco de Quito, Quito, Ecuador}
\affiliation{LPC, Universit\'e Blaise Pascal, CNRS/IN2P3, Clermont, France}
\affiliation{LPSC, Universit\'e Joseph Fourier Grenoble 1, CNRS/IN2P3, Institut National Polytechnique de Grenoble, Grenoble, France}
\affiliation{CPPM, Aix-Marseille Universit\'e, CNRS/IN2P3, Marseille, France}
\affiliation{LAL, Universit\'e Paris-Sud, CNRS/IN2P3, Orsay, France}
\affiliation{LPNHE, Universit\'es Paris VI and VII, CNRS/IN2P3, Paris, France}
\affiliation{CEA, Irfu, SPP, Saclay, France}
\affiliation{IPHC, Universit\'e de Strasbourg, CNRS/IN2P3, Strasbourg, France}
\affiliation{IPNL, Universit\'e Lyon 1, CNRS/IN2P3, Villeurbanne, France and Universit\'e de Lyon, Lyon, France}
\affiliation{III. Physikalisches Institut A, RWTH Aachen University, Aachen, Germany}
\affiliation{Physikalisches Institut, Universit{\"a}t Freiburg, Freiburg, Germany}
\affiliation{II. Physikalisches Institut, Georg-August-Universit{\"a}t G\"ottingen, G\"ottingen, Germany}
\affiliation{Institut f{\"u}r Physik, Universit{\"a}t Mainz, Mainz, Germany}
\affiliation{Ludwig-Maximilians-Universit{\"a}t M{\"u}nchen, M{\"u}nchen, Germany}
\affiliation{Fachbereich Physik, Bergische Universit{\"a}t Wuppertal, Wuppertal, Germany}
\affiliation{Panjab University, Chandigarh, India}
\affiliation{Delhi University, Delhi, India}
\affiliation{Tata Institute of Fundamental Research, Mumbai, India}
\affiliation{University College Dublin, Dublin, Ireland}
\affiliation{Korea Detector Laboratory, Korea University, Seoul, Korea}
\affiliation{CINVESTAV, Mexico City, Mexico}
\affiliation{Nikhef, Science Park, Amsterdam, the Netherlands}
\affiliation{Radboud University Nijmegen, Nijmegen, the Netherlands and Nikhef, Science Park, Amsterdam, the Netherlands}
\affiliation{Joint Institute for Nuclear Research, Dubna, Russia}
\affiliation{Institute for Theoretical and Experimental Physics, Moscow, Russia}
\affiliation{Moscow State University, Moscow, Russia}
\affiliation{Institute for High Energy Physics, Protvino, Russia}
\affiliation{Petersburg Nuclear Physics Institute, St. Petersburg, Russia}
\affiliation{Instituci\'{o} Catalana de Recerca i Estudis Avan\c{c}ats (ICREA) and Institut de F\'{i}sica d'Altes Energies (IFAE), Barcelona, Spain}
\affiliation{Stockholm University, Stockholm and Uppsala University, Uppsala, Sweden}
\affiliation{Lancaster University, Lancaster LA1 4YB, United Kingdom}
\affiliation{Imperial College London, London SW7 2AZ, United Kingdom}
\affiliation{The University of Manchester, Manchester M13 9PL, United Kingdom}
\affiliation{University of Arizona, Tucson, Arizona 85721, USA}
\affiliation{University of California Riverside, Riverside, California 92521, USA}
\affiliation{Florida State University, Tallahassee, Florida 32306, USA}
\affiliation{Fermi National Accelerator Laboratory, Batavia, Illinois 60510, USA}
\affiliation{University of Illinois at Chicago, Chicago, Illinois 60607, USA}
\affiliation{Northern Illinois University, DeKalb, Illinois 60115, USA}
\affiliation{Northwestern University, Evanston, Illinois 60208, USA}
\affiliation{Indiana University, Bloomington, Indiana 47405, USA}
\affiliation{Purdue University Calumet, Hammond, Indiana 46323, USA}
\affiliation{University of Notre Dame, Notre Dame, Indiana 46556, USA}
\affiliation{Iowa State University, Ames, Iowa 50011, USA}
\affiliation{University of Kansas, Lawrence, Kansas 66045, USA}
\affiliation{Kansas State University, Manhattan, Kansas 66506, USA}
\affiliation{Louisiana Tech University, Ruston, Louisiana 71272, USA}
\affiliation{Boston University, Boston, Massachusetts 02215, USA}
\affiliation{Northeastern University, Boston, Massachusetts 02115, USA}
\affiliation{University of Michigan, Ann Arbor, Michigan 48109, USA}
\affiliation{Michigan State University, East Lansing, Michigan 48824, USA}
\affiliation{University of Mississippi, University, Mississippi 38677, USA}
\affiliation{University of Nebraska, Lincoln, Nebraska 68588, USA}
\affiliation{Rutgers University, Piscataway, New Jersey 08855, USA}
\affiliation{Princeton University, Princeton, New Jersey 08544, USA}
\affiliation{State University of New York, Buffalo, New York 14260, USA}
\affiliation{Columbia University, New York, New York 10027, USA}
\affiliation{University of Rochester, Rochester, New York 14627, USA}
\affiliation{State University of New York, Stony Brook, New York 11794, USA}
\affiliation{Brookhaven National Laboratory, Upton, New York 11973, USA}
\affiliation{Langston University, Langston, Oklahoma 73050, USA}
\affiliation{University of Oklahoma, Norman, Oklahoma 73019, USA}
\affiliation{Oklahoma State University, Stillwater, Oklahoma 74078, USA}
\affiliation{Brown University, Providence, Rhode Island 02912, USA}
\affiliation{University of Texas, Arlington, Texas 76019, USA}
\affiliation{Southern Methodist University, Dallas, Texas 75275, USA}
\affiliation{Rice University, Houston, Texas 77005, USA}
\affiliation{University of Virginia, Charlottesville, Virginia 22901, USA}
\affiliation{University of Washington, Seattle, Washington 98195, USA}
\author{V.M.~Abazov} \affiliation{Joint Institute for Nuclear Research, Dubna, Russia}
\author{B.~Abbott} \affiliation{University of Oklahoma, Norman, Oklahoma 73019, USA}
\author{B.S.~Acharya} \affiliation{Tata Institute of Fundamental Research, Mumbai, India}
\author{M.~Adams} \affiliation{University of Illinois at Chicago, Chicago, Illinois 60607, USA}
\author{T.~Adams} \affiliation{Florida State University, Tallahassee, Florida 32306, USA}
\author{G.D.~Alexeev} \affiliation{Joint Institute for Nuclear Research, Dubna, Russia}
\author{G.~Alkhazov} \affiliation{Petersburg Nuclear Physics Institute, St. Petersburg, Russia}
\author{A.~Alton$^{a}$} \affiliation{University of Michigan, Ann Arbor, Michigan 48109, USA}
\author{G.~Alverson} \affiliation{Northeastern University, Boston, Massachusetts 02115, USA}
\author{G.A.~Alves} \affiliation{LAFEX, Centro Brasileiro de Pesquisas F{\'\i}sicas, Rio de Janeiro, Brazil}
% ...
\collaboration{The D0 Collaboration\footnote{with visitors from
$^{a}$Augustana College, Sioux Falls, SD, USA,
$^{b}$The University of Liverpool, Liverpool, UK,
$^{c}$SLAC, Menlo Park, CA, USA,
$^{d}$University College London, London, UK,
$^{e}$Centro de Investigacion en Computacion - IPN, Mexico City, Mexico,
$^{f}$ECFM, Universidad Autonoma de Sinaloa, Culiac\'an, Mexico,
and 
$^{g}$Universit{\"a}t Bern, Bern, Switzerland.
$^{\ddag}$Deceased.
}} \noaffiliation
\vskip 0.25cm
       
\date{July 22, 2011}
\begin{abstract}
We present a search for the standard model Higgs boson and a fermiophobic Higgs boson 
in the diphoton final states based on 8.2 fb$^{-1}$ of  $p\bar{p}$ collisions
at $\sqrt{s} = 1.96$ TeV collected with
the D0 detector at the Fermilab Tevatron Collider. 
No excess of data above background predictions
is observed and upper limits at the 95\% C.L.  on the cross section 
multiplied by the branching fraction are set which are the most restrictive to date. 
A fermiophobic Higgs boson with a
mass below 112.9 GeV is excluded at the 95\% C.L.
\end{abstract}
\pacs{14.80.Bn, 13.85.Rm, 14.80.Ec, 12.60.Fr}
\maketitle
In the standard model (SM), the Higgs boson ($H$) is the last undiscovered particle that provides 
crucial insights on the spontaneous breaking of the electroweak symmetry and the generation of 
mass of the weak gauge bosons and fermions.
The constraints from the direct searches at the CERN $e^+e^-$ Collider (LEP) \cite{sm-1} and
from the measurement of precision electroweak observables~\cite{sm-2} result in a preferred range
for the SM Higgs boson mass of $114.4<M_{H}<185$~GeV at 95\% C.L.
Furthermore, the range $158< M_{H} <173$~GeV is excluded at 95\% C.L. 
by the direct searches at the Fermilab Tevatron $p\bar{p}$ Collider~\cite{sm-3}. 
These experimental constraints are derived assuming SM production and decay modes 
for the Higgs boson and can be substantially modified in case of significant departures from the SM.

At hadron colliders the dominant production mechanisms for a light SM Higgs boson are
gluon fusion (GF) ($gg\to H$), associated production 
with a $W$ or $Z$ boson ($q\bar{q}^\prime \to VH$, $V=W,Z$), and vector boson fusion (VBF) 
($VV\to H$). At the Tevatron the most sensitive SM Higgs boson searches rely on the 
$VH (H\to b\bar{b})$ process for  $M_H<125$~GeV and on $gg\to H\to W^+W^-$ for  $M_H>125$~GeV.
At CERN's Large Hadron Collider (LHC), the strategy at high $M_{H}$ ($>$ 140 GeV) is similar, 
while at low $M_{H}$ ($<$ 140 GeV) the $H\to \gamma\gamma$ decay mode 
becomes one of the most promising discovery channels, despite its small branching ratio of 
${\cal B} (H\to\gamma\gamma) \approx 0.2\%$ for $110<M_H<140$~GeV, owing to its clean experimental signature of 
a narrow resonance on top of a smoothly-falling background in the diphoton mass spectrum. 
Some of the most sensitive searches for the SM Higgs boson involve
the loop-mediated $ggH$ and/or $\gamma\gamma H$ vertices, which are
sensitive to new physics effects. For instance, the addition of a sequential fourth family of quarks
can substantially enhance the $ggH$ coupling, leading to an increase in the GF production rate,
while decreasing ${\cal B} (H\to b\bar{b})$~\cite{SM4}. Alternatively, other models of electroweak
symmetry breaking can involve suppressed couplings to some or all fermions~\cite{beyond-SM}, 
with the extreme case being the fermiophobic Higgs boson ($H_f$) model, where the GF production
mode is absent, decays into fermions are heavily suppressed, and 
${\cal B} (H_f \to\gamma\gamma)$ is significantly enhanced.
Therefore, Higgs boson searches in the $\gamma\gamma$ decay
mode can be a sensitive probe of new physics models where the Higgs boson may be difficult
to observe in other, a priori more promising, channels.

This Letter presents a search for a Higgs boson decaying into $\gamma\gamma$ using an inclusive
diphoton sample collected with the D0 detector in $p\bar{p}$ collisions at $\sqrt{s}=1.96$~TeV
at the Fermilab Tevatron Collider. In this search both the SM and the fermiophobic Higgs
boson models are considered. The most recent searches at the Tevatron for a SM Higgs boson~\cite{D0-SM} 
or a fermiophobic Higgs boson~\cite{TEV-FH} in the $\gamma\gamma$ mode analyzed the 
diphoton invariant mass spectrum in search for a narrow resonance. This analysis represents a significant step forward 
in sensitivity by increasing the dataset by nearly a factor of three, as well as by exploiting further kinematic
differences between signal and background through a multivariate analysis technique. 

The D0 detector is described in detail elsewhere~\cite{d0det}.
The subdetectors most relevant to this analysis are the central tracking system,
composed of a silicon microstrip tracker (SMT) and a central fiber tracker (CFT)
in a 2 T solenoidal magnetic field, the central preshower (CPS), and
the liquid-argon and uranium sampling calorimeter. The CPS is located immediately
before the inner layer of the calorimeter and is formed by one radiation length
of absorber followed by several layers of scintillating strips.
The calorimeter consists of three sections housed in separate cryostats: a central 
section covering up to $|\eta|\approx 1.1$~\cite{EtaCov} 
and two end calorimeters extending the coverage up to $|\eta|\approx 4.2$.
They are divided into electromagnetic (EM) and hadronic layers.
The EM section of the calorimeter is segmented into four
longitudinal layers with transverse segmentation of $\Delta\eta \times \Delta\phi = 0.1\times 0.1$~\cite{EtaCov},
except in the third layer (EM3), where it is $0.05\times 0.05$. The calorimeter
is well suited for a precise measurement of electron and photon energies, 
providing a resolution of $\approx 3.6\%$ at electron and photon energies of $\approx 50$~GeV. The data
used in this analysis were collected using triggers requiring at least two
clusters of energy in the EM calorimeter and correspond to an integrated luminosity
of $8.2$~fb$^{-1}$~\cite{d0lumi}.

Events are selected by requiring at least two photon candidates with transverse momentum
$p_T>25$ GeV in the central region of the calorimeter ($|\eta|<1.1$), for which the trigger 
requirements are close to $100\%$ efficient.  Photon candidates are selected from EM clusters 
reconstructed with a simple cone algorithm with radius ${\cal R}=\sqrt{(\Delta \eta)^2 + (\Delta \phi)^2}=0.2$ 
that satisfy the following requirements:
(i) at least $95\%$ of the cluster energy is deposited in the
EM calorimeter; (ii) the calorimeter isolation variable
$I = [E_{\text{tot}}(0.4)-E_{\text{EM}}(0.2)]/E_{\text{EM}}(0.2)$ is
less than 0.1, where $E_{\text{tot}}(0.4)$ is the
total energy in a cone of radius ${\cal R}=0.4$ and
$E_{\text{EM}}(0.2)$ is the EM energy in a cone of radius ${\cal R}=0.2$;
(iii) the energy-weighted cluster width
in EM3 is consistent with an EM shower;
(iv) the scalar sum of the $p_T$ of all tracks originating from the
primary $p\bar{p}$ interaction vertex in an annulus of $0.05<{\cal R}<0.4$ around the cluster 
is less than 2 GeV;
(v) the EM cluster is not spatially matched to tracker activity, either to a reconstructed track, 
or to a set of hits in the SMT and CFT consistent with that of an electron or positron trajectory \cite{HOR}; and 
(vi) the output of a photon neural network ($O_{NN}$)~\cite{D0-SM, EPAPS}, combining information
from a set of variables that are sensitive to differences between photons and jets 
in the tracker, the calorimeter and the CPS, is larger than 0.1.
The requirement (vi) rejects approximately 40\% of the misidentified jets, while keeping $> 98\%$ of real photons.
Finally, additional kinematic selections are applied in order to select a signal-enriched sample.
The diphoton invariant mass, $\mgg$, computed from the two highest $p_T$ photons in an event,
is required to be larger than 60~GeV.
The azimuthal angle between the two
photons, $\dphigg$, is required to be larger than 0.5,
which reduces the background from events where both photons originate
from fragmentation, a process that is not well modeled in the simulation,
while keeping $> 99\%$ of the Higgs boson signal. 

\begin{table*}
\caption{\label{event-tab} \small Signal, backgrounds and data yields for $M_{H} = 100$ GeV to 150 GeV in 10 GeV intervals
within the [$M_{H}$ - 30 GeV, $M_{H}$ + 30 GeV] mass window. The background yields result from a fit to the
data. The uncertainties include both statistical and systematic contributions and take
into account correlations among processes.
The uncertainty on the total background is smaller than the sum in 
quadrature of the uncertainties in the individual background sources
due to the anti-correlation resulting from the fit.
}
\centering
\begin{tabular}{lcccccc}
\hline \hline
$M_H$ (GeV) &100 & 110 & 120 & 130 & 140 & 150  \\ \hline

$\gamma\gamma$ (DDP)    & 6415 $\pm$ 395 &4031 $\pm$ 286  &2779 $\pm$ 188  &1849 $\pm$ 139  &1355 $\pm$ 99   & 1026 $\pm$ 75\\
$\gamma j + jj$         & 5727 $\pm$ 352 &3819 $\pm$ 252  &2265 $\pm$ 178  &1506 $\pm$ 120  &964 $\pm$ 87   &  641 $\pm$ 63\\
$Z/\gamma^{*} \rightarrow e^+e^-$
                        &  599 $\pm$ 91  & 517 $\pm$ 81   & 361 $\pm$ 55   & 141 $\pm$ 23   &65 $\pm$ 12    &   34 $\pm$ 7\\
\hline
Total background       & 12741 $\pm$ 160 &8367 $\pm$ 134  &5405 $\pm$ 95  &3496 $\pm$ 77   &2384 $\pm$ 57   & 1701 $\pm$ 48\\

Data & 12746 & 8380 & 5406 & 3500 & 2383 & 1696\\
\hline
$H$ boson signal       &5.9 $\pm$ 0.8 & 5.8 $\pm$ 0.8 &  5.3 $\pm$ 0.7 & 4.2 $\pm$ 0.6 & 2.9 $\pm$ 0.4 & 1.7 $\pm$ 0.2 \\
$H_{f}$ boson signal &149.7 $\pm$ 13.2&39.4 $\pm$ 3.5 & 11.9 $\pm$ 1.0 & 4.4 $\pm$ 0.4 & 1.8 $\pm$ 0.2 & 0.7 $\pm$ 0.1 \\

\hline \hline
\end{tabular}
\end{table*}

The selected data sample is contaminated by backgrounds of instrumental
origin such as $\gamma$+jet ($\gamma j$), dijet ($jj$) and $Z/\gamma^*\rightarrow e^{+}e^{-}$ (ZDY) production,
with jets or electrons misidentified as photons, as well as a background
from direct $\gamma\gamma$ production (DDP) where two isolated photons are produced.
The normalization and shape of the $\gamma j$ and $jj$ backgrounds, as well as
the overall normalization of the DDP background, are estimated from data.
The Monte Carlo (MC) simulation is used to model the normalization and shape of
the signal and ZDY background, as well as the shape of the DDP background.
The MC samples used in this analysis are generated using {\sc pythia}~\cite{pythia}
(for signal and ZDY) or {\sc sherpa}~\cite{sherpa} (for DDP) with CTEQ6L1~\cite{cteq} parton 
distribution functions (PDFs), followed by a {\sc geant}-based~\cite{geant}
simulation of the D0 detector. Events from randomly selected beam crossings are overlaid on the simulated events
to better model contributions from additional $p\bar{p}$ interactions and detector noise.
The same reconstruction algorithms are used as on the data.
Signal samples are generated separately for the GF, VH and VBF processes and
normalized using the next-to-next-to-leading order (NNLO) 
plus next-to-next-to-leading-logarithm (NNLL) theoretical cross sections for GF 
and NNLO for VH and VBF processes~\cite{ggHxsect,VHxsect,VBFxsect}.
The Higgs boson's branching ratio predictions are from {\sc hdecay}~\cite{hdecay}.
The ZDY background estimate from MC is normalized to the NNLO cross section~\cite{Z-Xsection}. 

The $\gamma j$ and $jj$ yields are estimated with data~\cite{bkg-subtract}.
Following the final selection,  a tightened $O_{NN}$ requirement ($O_{NN}>0.75$) is used
to classify the events into four categories: (i) both photons, (ii) only the highest $p_T$ (leading) photon, 
(iii) only the second highest $p_T$ (trailing) photon, or (iv) neither of the two photons, satisfy this requirement. 
The corresponding numbers of events, after subtracting the ZDY contribution, are denoted as (i) $N_{pp}$, (ii) $N_{pf}$, (iii) $N_{fp}$ and (iv) $N_{ff}$.
The different efficiency of the $O_{NN}>0.75$ requirement for photons ($\epsilon_{\gamma}$) and jets ($\epsilon_{\text {jet}}$) is used to estimate the sample composition by solving a linear system of equations:
$$(N_{pp}, N_{pf}, N_{fp}, N_{ff})^T = {\cal E} \times (N_{\gamma \gamma}, N_{\gamma j}, N_{j \gamma}, N_{jj})^T ,$$
where $N_{\gamma\gamma}$ ($N_{jj}$) is the number of $\gamma\gamma$ ($jj$) events and $N_{\gamma j}$ ($N_{j \gamma}$)
is the number of $\gamma j$ events with the leading (trailing) cluster as the photon.
The $4\times4$ matrix ${\cal E}$ contains the efficiency terms
$\epsilon_{\gamma}$ and $\epsilon_{\text {jet}}$, parameterized as a function of $|\eta|$ for each photon candidate and
estimated in photon and jet MC samples.  We validate $\epsilon_{\gamma}$ with 
data of radiated photon from charged leptons in $Z$ boson decays ($Z \rightarrow l^+l^-\gamma, l=e,\mu$)
and $\epsilon_{\text {jet}}$ with jet data~\cite{diphoton-Xsection}. 
The DDP normalization is determined from a fit to the final discriminant distribution used for hypothesis testing, exploiting the
difference in shape between signal and background
in each $M_H$ search region. 
For each $M_H$ hypothesis (between 100 and 150~GeV in steps
of 2.5 GeV), the search region is defined to be
$M_{H} \pm 30$~GeV. 
The shape of the DDP background is obtained from {\sc sherpa}~\cite{sherpa}, while the shapes of the $\gamma j$ and $jj$ backgrounds are obtained from
independent data control samples selected by requiring exactly one photon or both photon candidates to satisfy $O_{NN}<0.1$, respectively.
Table~\ref{event-tab} shows the numbers of data events, expected background, and the expected
$H$ boson and $H_f$ boson signals in six of the search regions
resulting from a fit described later in this Letter. 
The estimated background composition is $\approx 53\%$ from DDP, $\approx 44\%$ from $\gamma j$+$jj$ and $\approx 3\%$ from ZDY.

\begin{figure*}[t]
 \centering
 \epsfig{file=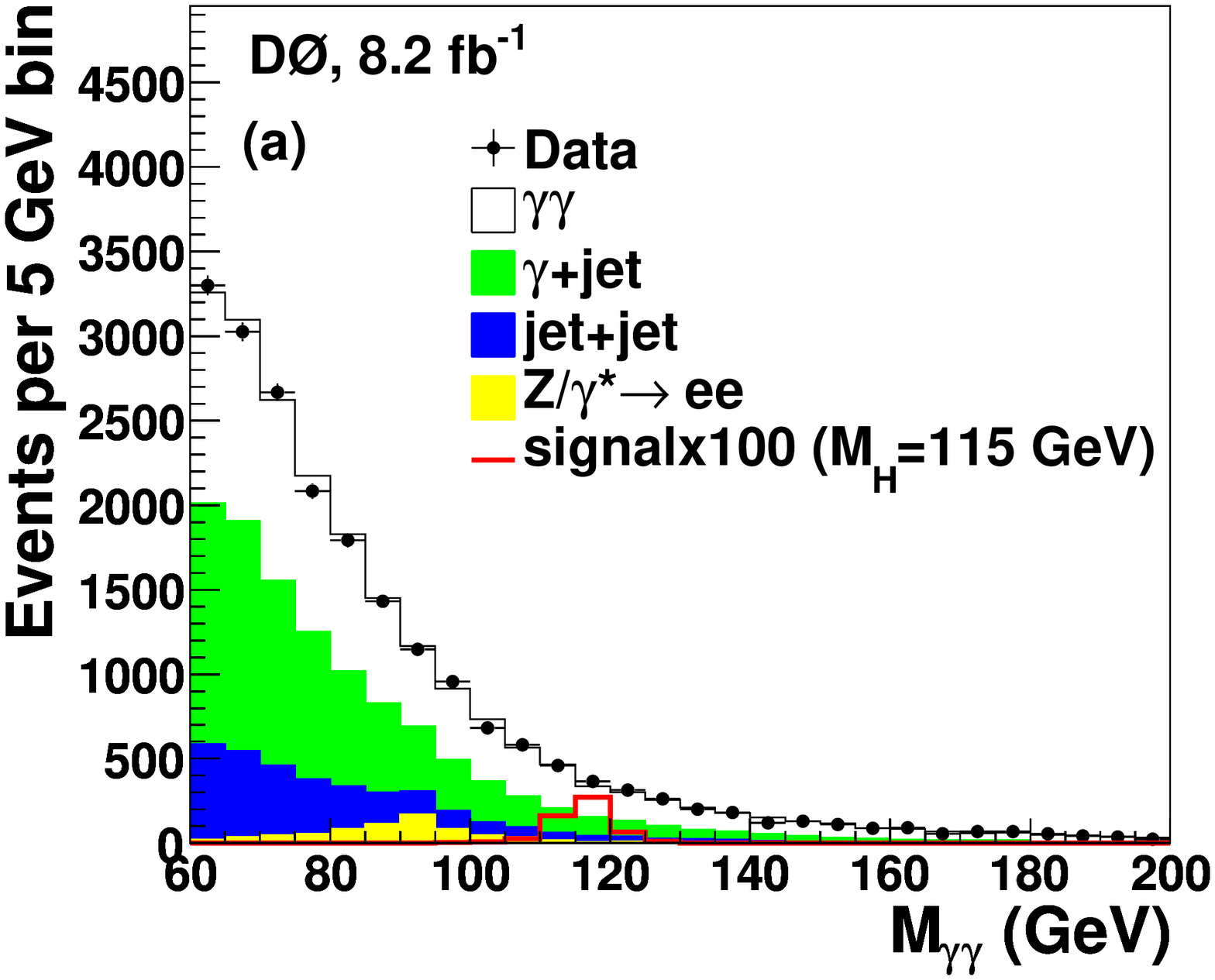,scale=0.29}
 \epsfig{file=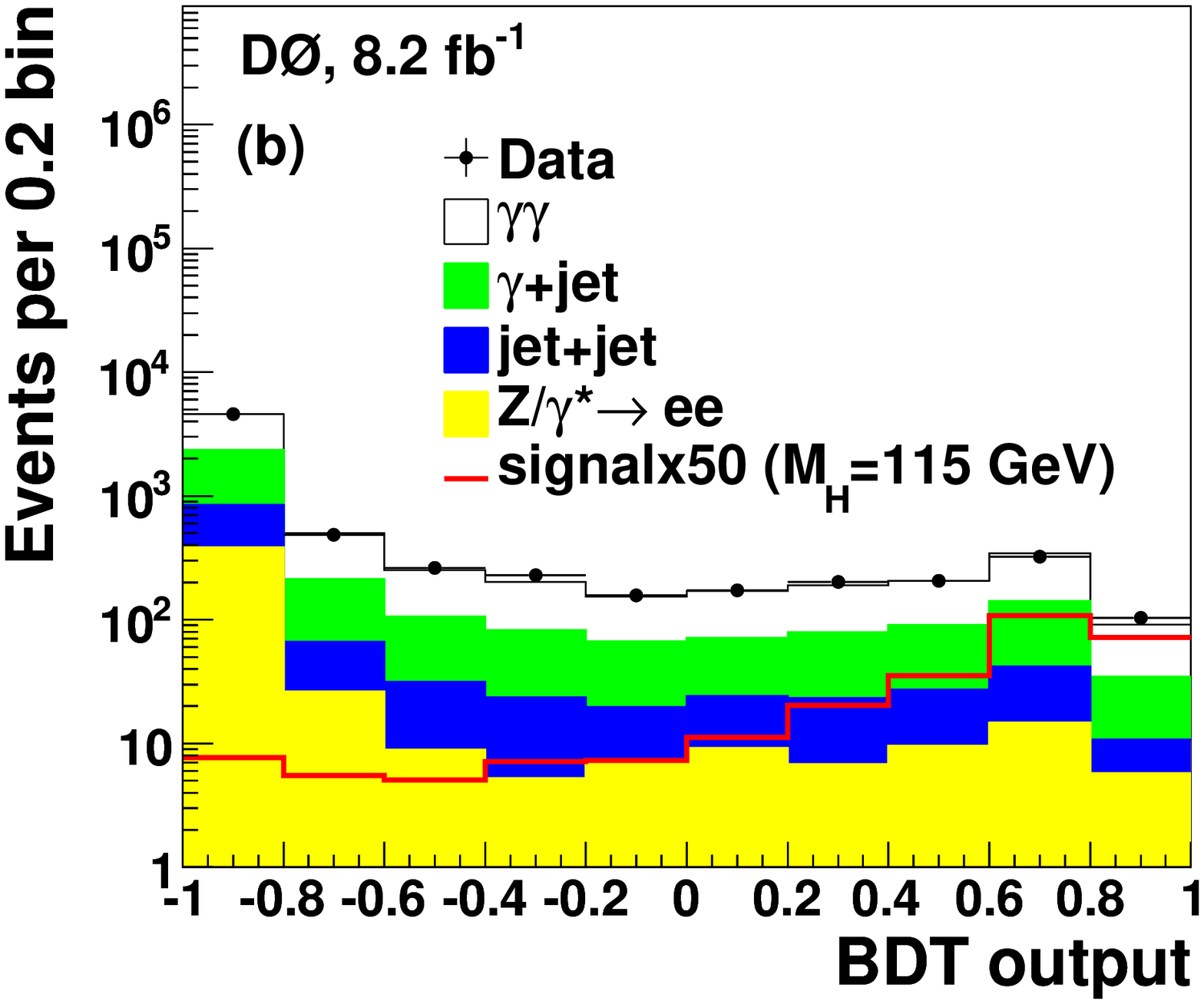,scale=0.29}
 \epsfig{file=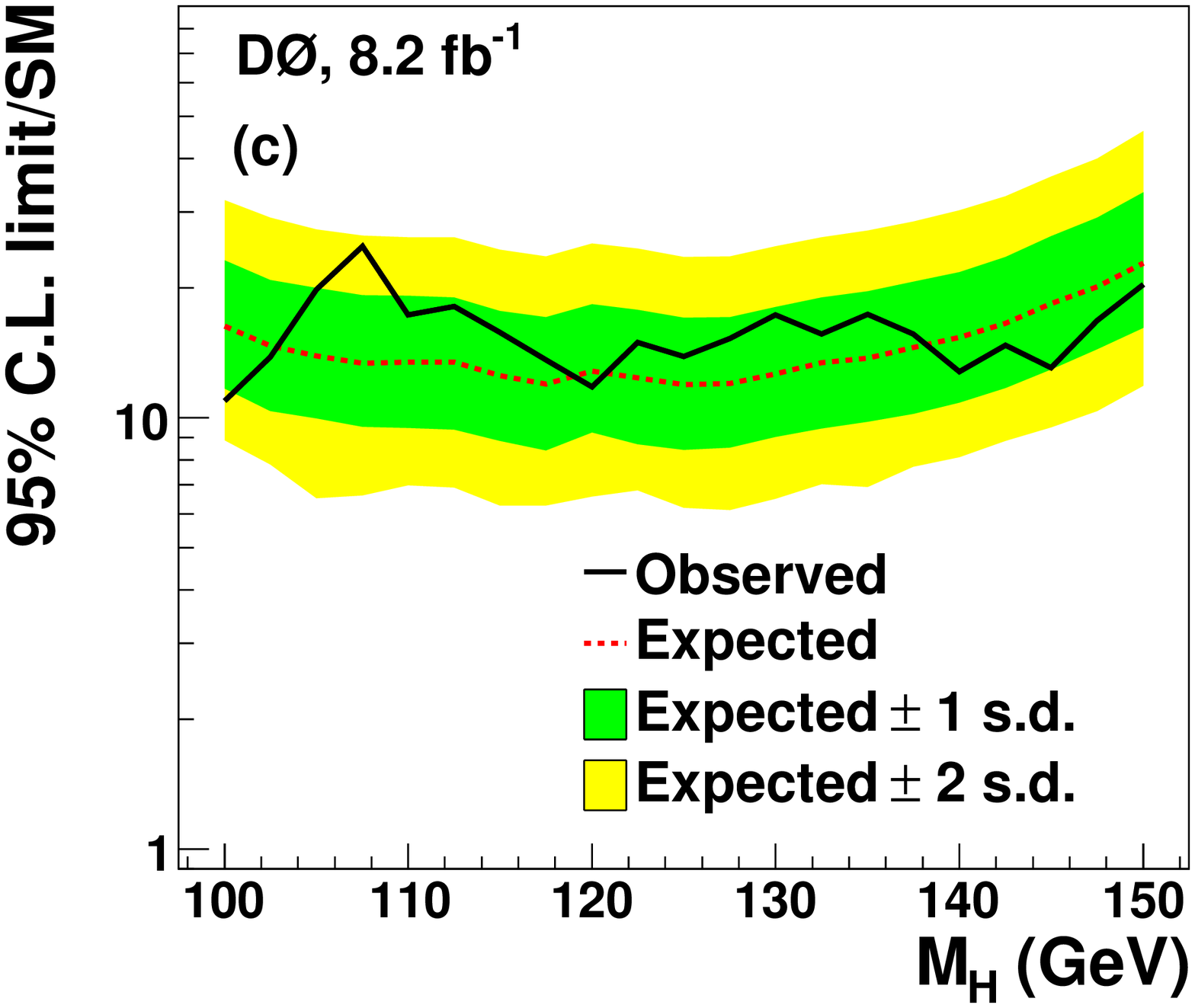,scale=0.29} 
   \caption{\small (color online). 
    (a) $\mgg$ and (b) BDT output distributions for $M_H=115$~GeV
after the final selection comparing data to the background prediction. The expected $H$ boson signal multiplied
by a factor of 50 is also shown.
(c) Observed and expected 95\%~C.L. upper limits on $\sigma \times {\cal B}$ relative to the SM prediction as a function of $M_H$.
The bands correspond to the $\pm$~1 and $\pm$~2 standard deviations (s.d.) around the expected limit
under the background-only hypothesis.}
  \label{result1}
\end{figure*}

To improve the sensitivity of the search, a total of five well-modeled kinematic variables are
used to discriminate between signal and background: $\mgg$, $\dphigg$,
the transverse momentum of the diphoton system ($p_T^{\gamma\gamma}$), and the transverse momenta of the
leading and trailing photons ($p_T^{1}$, $p_T^{2}$). Figure~\ref{result1}(a) shows a comparison of the $\mgg$ distribution 
between data and the background prediction. Comparisons for other kinematic distributions can be found in
Ref.~\cite{EPAPS}. A boosted-decision-tree (BDT) technique~\cite{bdt} is used to build a
single discriminant variable combining the information from the above five variables. 
A different BDT is trained for each $M_H$ hypothesis, separately for the SM and the fermiophobic Higgs boson models. 
In each model, the training
is performed to discriminate between the sum of all relevant signals
and the sum of all backgrounds. Figure~\ref{result1}(b) shows a comparison of the BDT output distribution
between data and background prediction corresponding to the SM for $M_H=115$~GeV.

\begin{figure}[htbp]
 \centering
  \epsfig{file=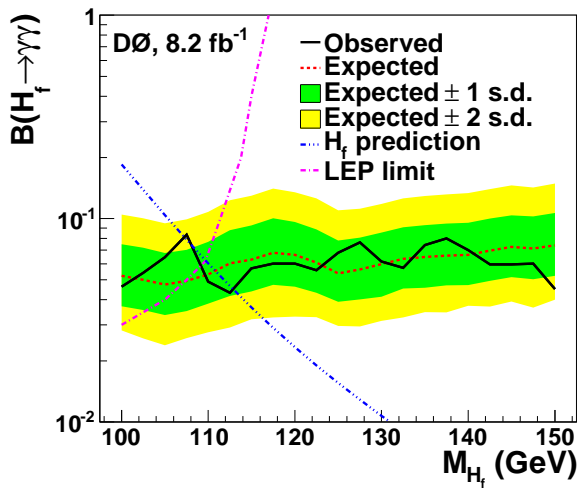,scale=0.4}
   \caption{\small (color online). Observed and expected 95\%~C.L. upper limits on ${\cal B} (H_f \to \gamma\gamma)$ as a function of $M_{H_{f}}$. 
    The definition of the bands are the same as in Fig.~\ref{result1}(c).
    The blue line represents the branching ratio predictions from {\sc hdecay}~\cite{hdecay}. 
    Also displayed is the exclusion region obtained by the LEP Collaborations~\cite{LEP-FH}.}
  \label{result2}
\end{figure}

Systematic uncertainties affecting the normalization and shape of the BDT output distribution
are estimated for both signal and backgrounds, taking into account correlations.
The sources of systematic uncertainties affecting the signal and ZDY background normalizations include the 
integrated luminosity (6.1\%), photon identification efficiency for signal (3.9\%) or electron 
misidentification rate for ZDY (12.7\%) and theoretical cross sections (including scale and 
PDF uncertainties) for signal (GF (14.1\%), VH (6.2\%) and VBF (4.9\%)) and ZDY (3.9\%) production.
The normalization uncertainty affecting the $\gamma j$+$jj$ prediction is $8.4\%$. This uncertainty results from propagating
the uncertainty on the $O_{NN}>0.75$ efficiency for photons (1.5\%) and jets (10\%) 
and also affects the shape of the $\gamma j$+$jj$ background at the 1\%-2\% level through
changes in the fractions of $\gamma j$ and $jj$.  Additional systematic uncertainties
affecting the differential distributions of data and MC include the relative photon energy scale (1\%-5\% for signal, 1\%-4\% for DDP),
DDP modeling  (1\%-10\%) and Higgs boson $p_T$ modeling in GF (1\%-5\%). Modeling uncertainties 
are obtained by changing the factorization and renormalization scales by a factor of two with respect to the 
nominal choice. 

No evidence for a signal, either in the SM or in the fermiophobic interpretations,
is found, and the BDT discriminants are used to derive upper limits  on the production
cross section multiplied by the branching ratio for $H\to\gamma\gamma$ ($\sigma \times {\cal B}$)
as a function of $M_H$. Limits are calculated at the 95\% C.L. with the CL$_{\text S}$ modified 
frequentist approach using a log-likelihood ratio of the signal-plus-background
(S+B) hypothesis to the background-only (B) hypothesis  \cite{CLs-1}. 
Systematic uncertainties are taken into account by convoluting the Poisson probability
distributions for signal and background with the corresponding Gaussian distributions.
The individual likelihoods are maximized with respect to the DDP background normalization 
as well as parameters that describe the systematic uncertainties~\cite{CLs-2}.
This fit allows the determination of the normalization for the DDP background from data 
and significantly reduces the impact of systematic uncertainties on the overall sensitivity.

The resulting upper limits on $\sigma \times {\cal B}$ relative to the SM prediction 
as a function of $M_H$ are shown in Fig.~\ref{result1}(c), representing the most constraining results 
for a SM Higgs boson decaying into photons. Upper limits on ${\cal B} (H_f \to \gamma\gamma)$ 
as a function of $M_{H_{f}}$ are presented in Fig.~\ref{result2} and compared to the LEP result~\cite{LEP-FH}. 
The sensitivity is improved by about a factor of two relative to previous
searches at the Tevatron~\cite{TEV-FH},
yielding the most stringent limits on a fermiophobic Higgs boson
of $M_{H_f}>112.9$~ GeV at 95\%~C.L.

Supplementary material is provided in~\cite{EPAPS}.

% acknowledgement.tex                             2 June 2011
%
We thank the staffs at Fermilab and collaborating institutions,
and acknowledge support from the
DOE and NSF (USA);
CEA and CNRS/IN2P3 (France);
FASI, Rosatom and RFBR (Russia);
CNPq, FAPERJ, FAPESP and FUNDUNESP (Brazil);
DAE and DST (India);
Colciencias (Colombia);
CONACyT (Mexico);
KRF and KOSEF (Korea);
CONICET and UBACyT (Argentina);
FOM (The Netherlands);
STFC and the Royal Society (United Kingdom);
MSMT and GACR (Czech Republic);
CRC Program and NSERC (Canada);
BMBF and DFG (Germany);
SFI (Ireland);
The Swedish Research Council (Sweden);
and
CAS and CNSF (China).

\newpage

\begin{table*}
\begin{center}
{\Large{\bf Supplementary material}}
\end{center}
\end{table*}

\begin{figure*}[t]
 \centering
  \epsfig{file=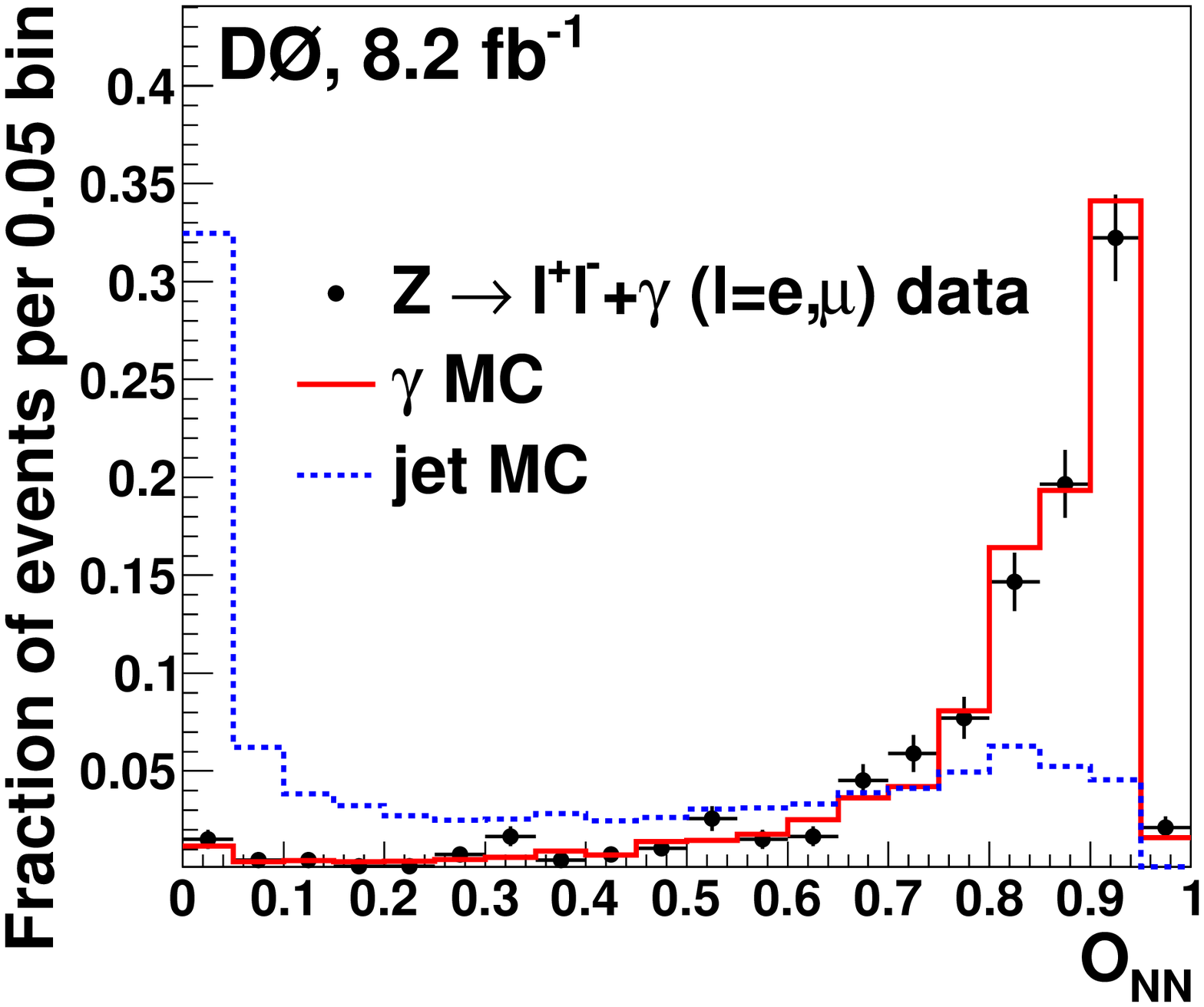,scale=0.4}
   \caption{\small Normalized $O_{NN}$ spectrum for photons from radiative $Z$ boson decays, 
   photons from diphoton MC and jets from dijet MC. The $O_{NN}$ distribution has strong discriminanting power
   between photons and jets, and the performances in photon data and MC are consistent. }
  \label{photonANN}
\end{figure*}

\begin{figure*}[t]
 \centering
\includegraphics[scale=0.4]{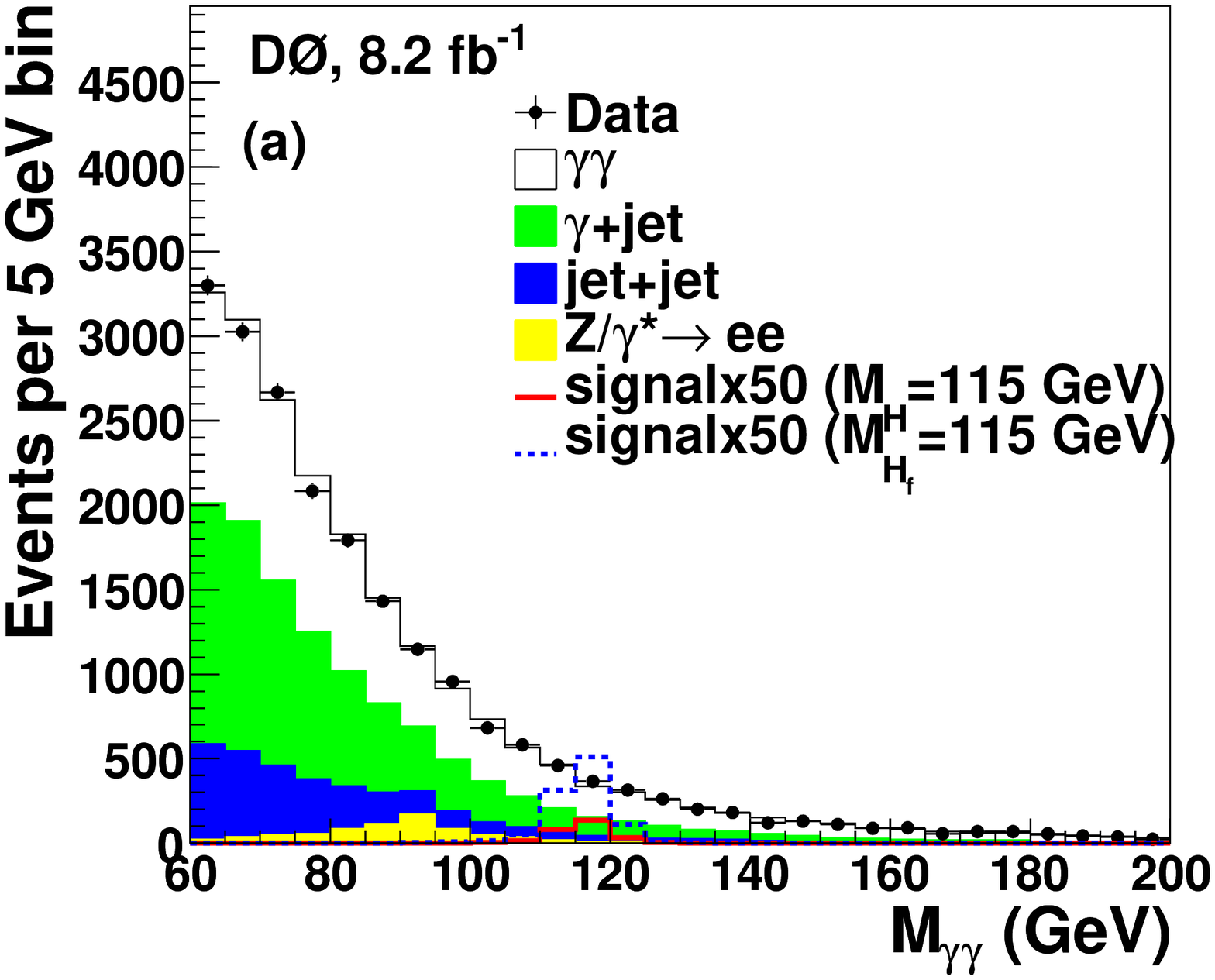}
\includegraphics[scale=0.4]{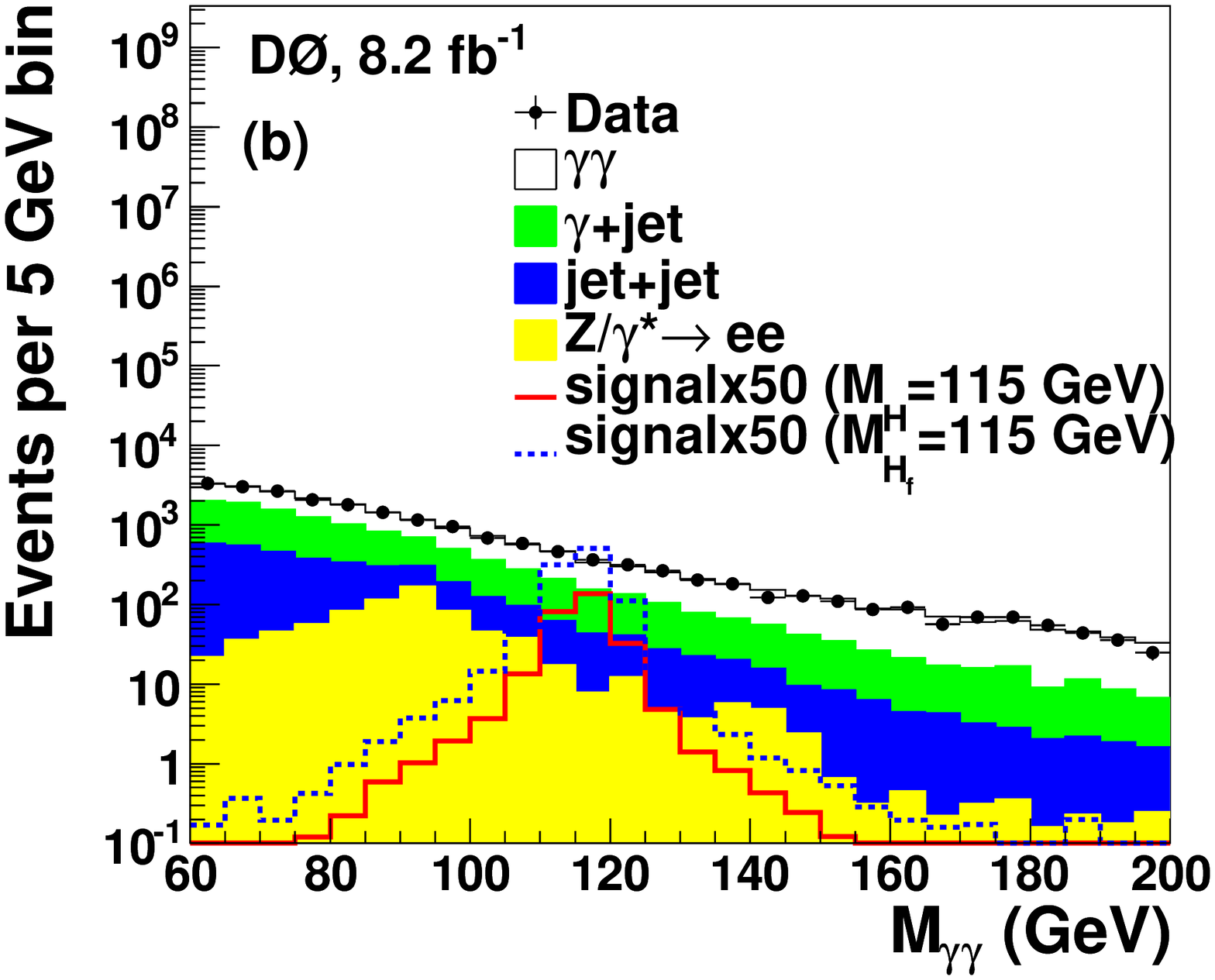} \\
\includegraphics[scale=0.4]{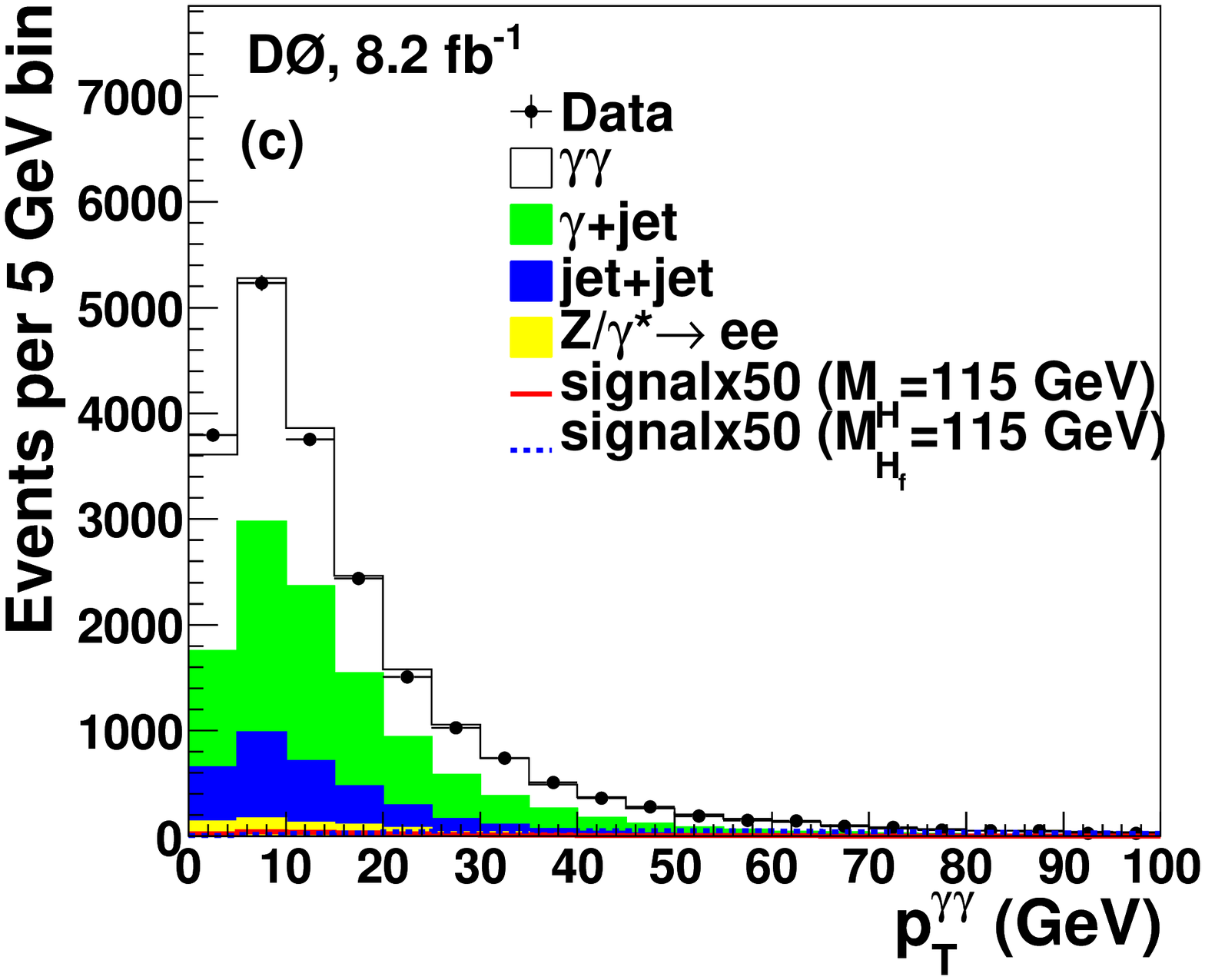}
\includegraphics[scale=0.4]{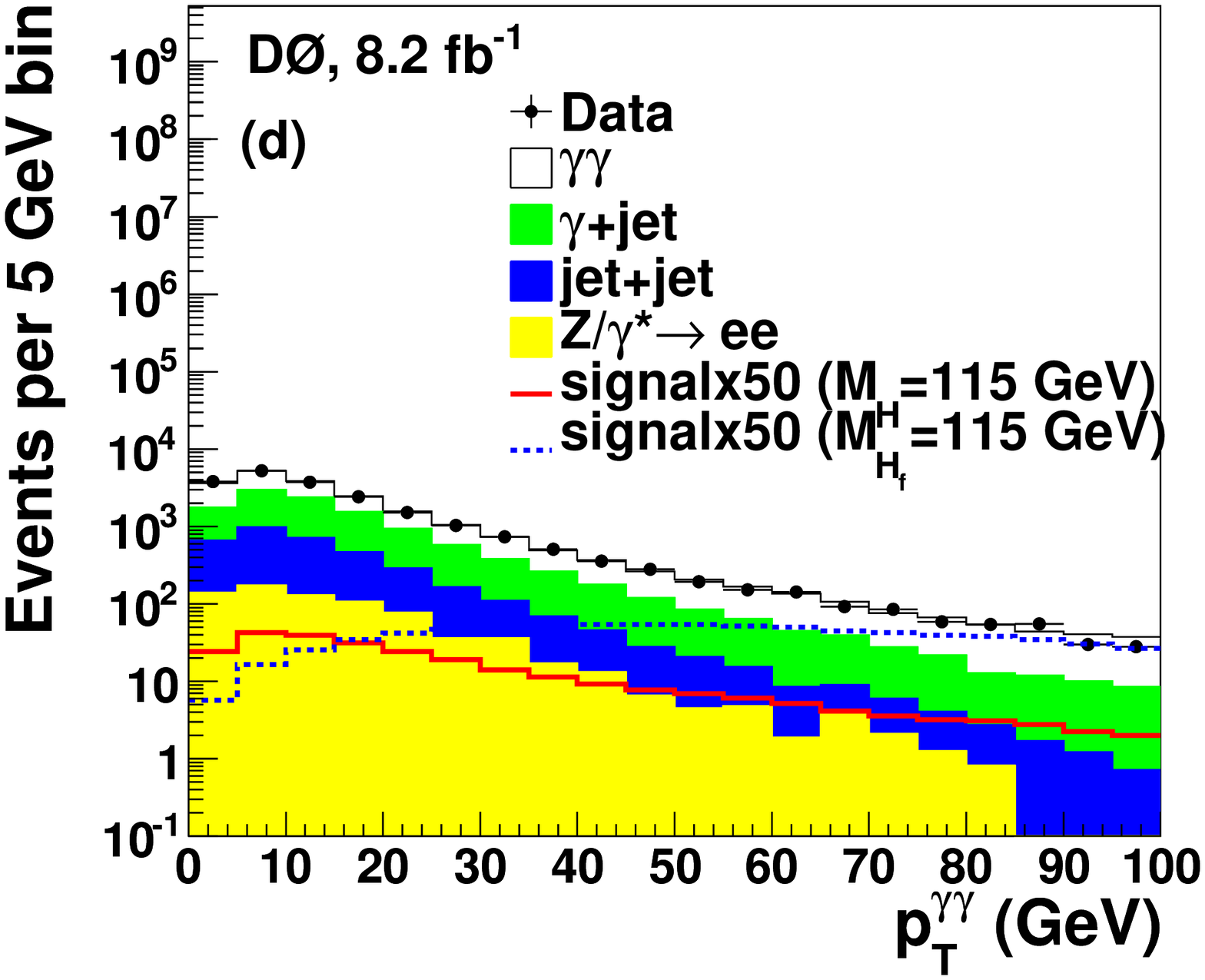} \\
\includegraphics[scale=0.4]{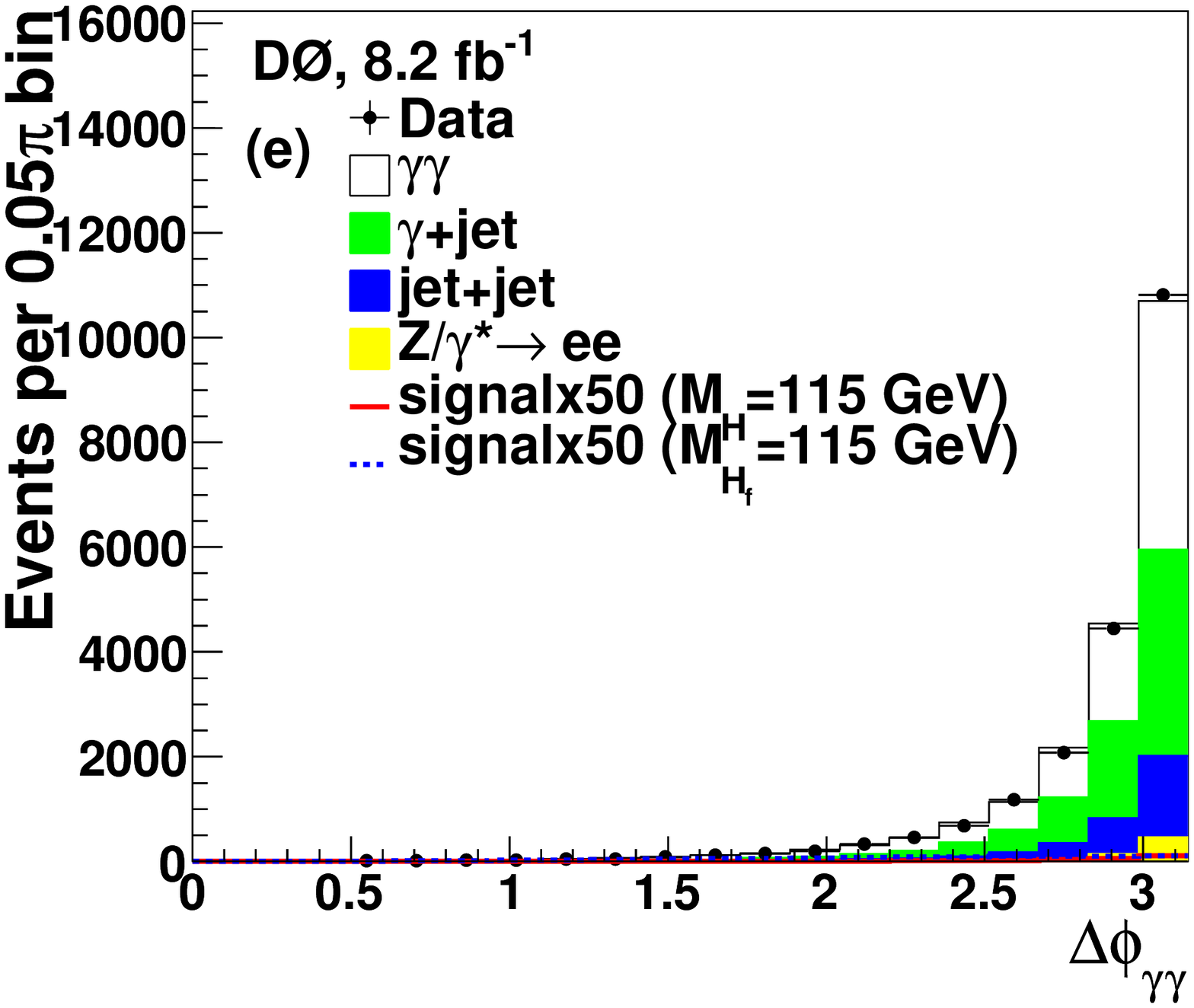}
\includegraphics[scale=0.4]{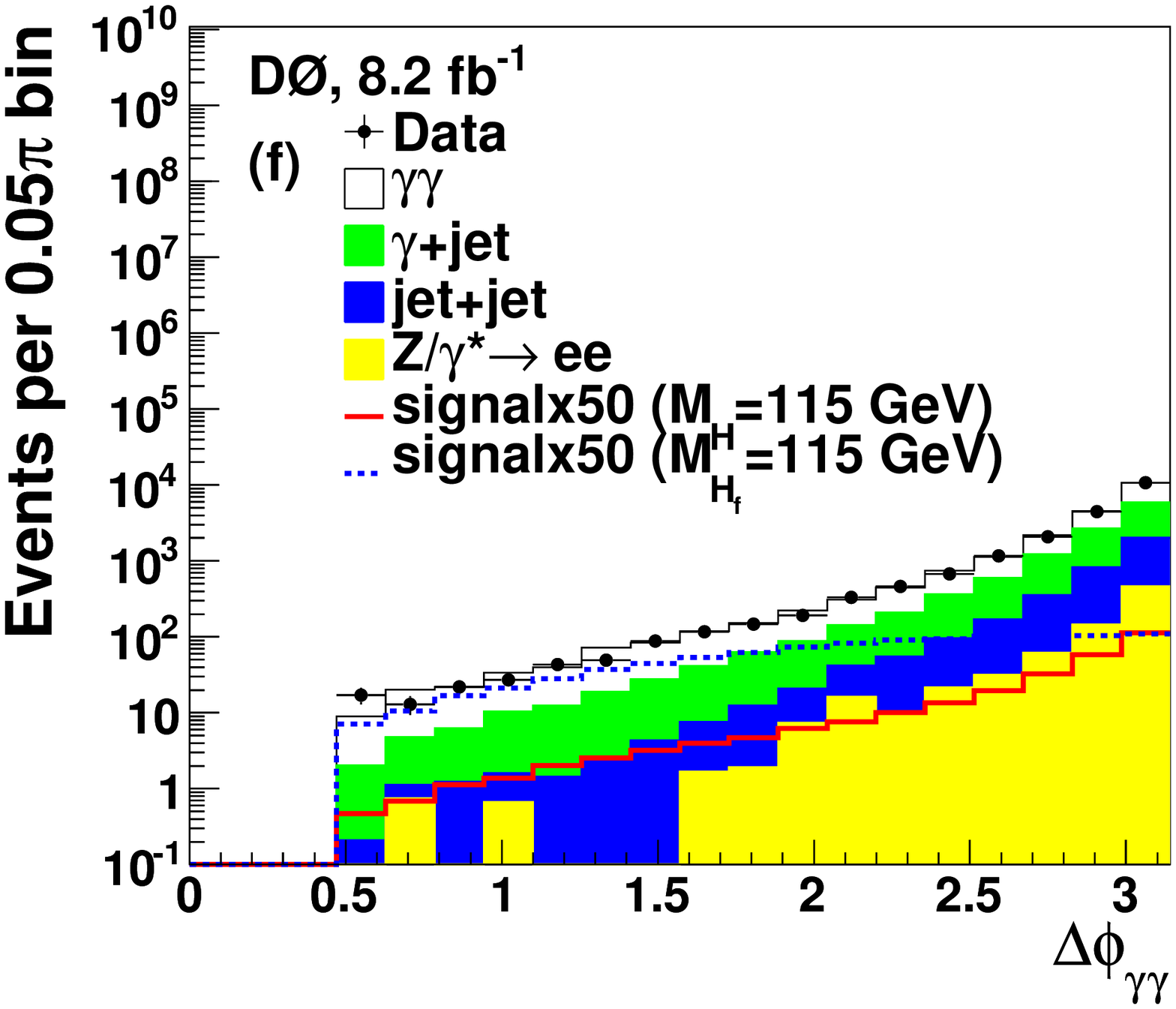}
   \caption{\small Variables used as input to the BDT: (a, b) $\mgg$, (c, d) $p_T^{\gamma\gamma}$ and (e, f) $\dphigg$,
   in linear scale (left column) and logarithmic scale (right column).
   Data are compared to the background prediction. Also shown
   is the expected signal for a SM Higgs boson ($M_H=115$~GeV) and a fermiophobic Higgs boson ($M_{H_{f}}=115$~GeV) 
   both multiplied by a factor of 50. For the fermiophobic Higgs boson, the GF production is absent and the diphoton system is on 
   average more boosted.}   
  \label{plots1}
\end{figure*}

\begin{figure*}[t]
 \centering
 \includegraphics[scale=0.4]{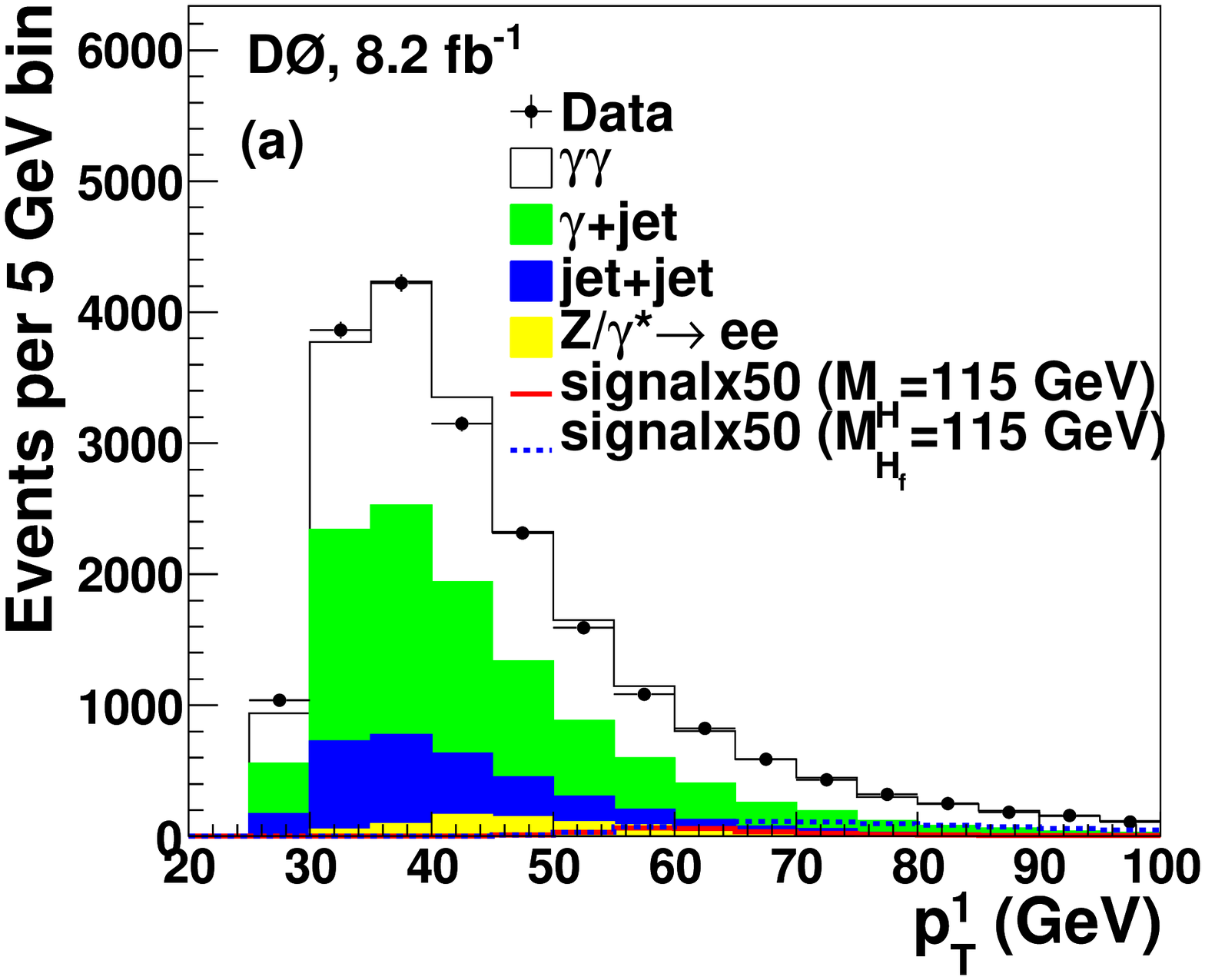}
 \includegraphics[scale=0.4]{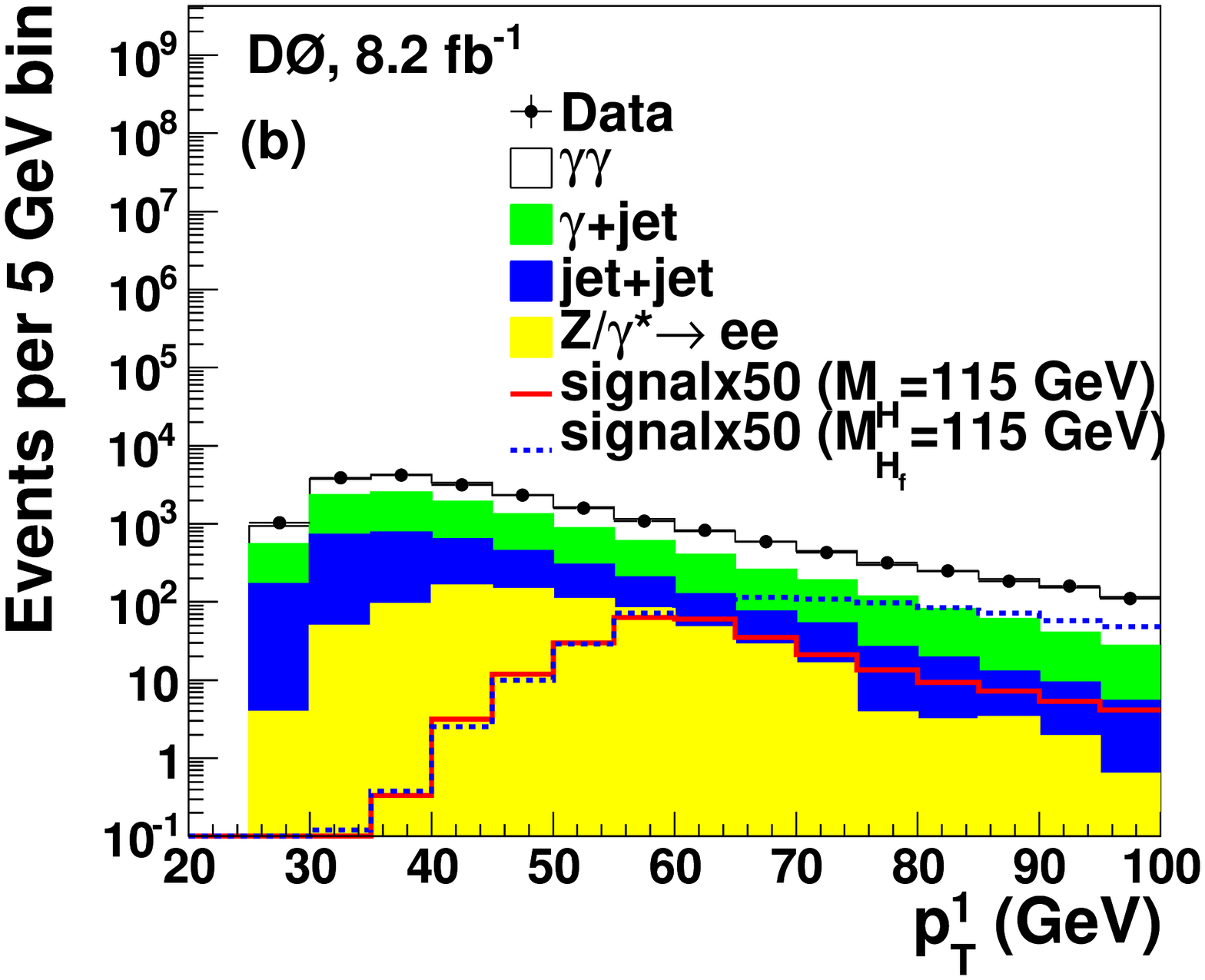} \\
 \includegraphics[scale=0.4]{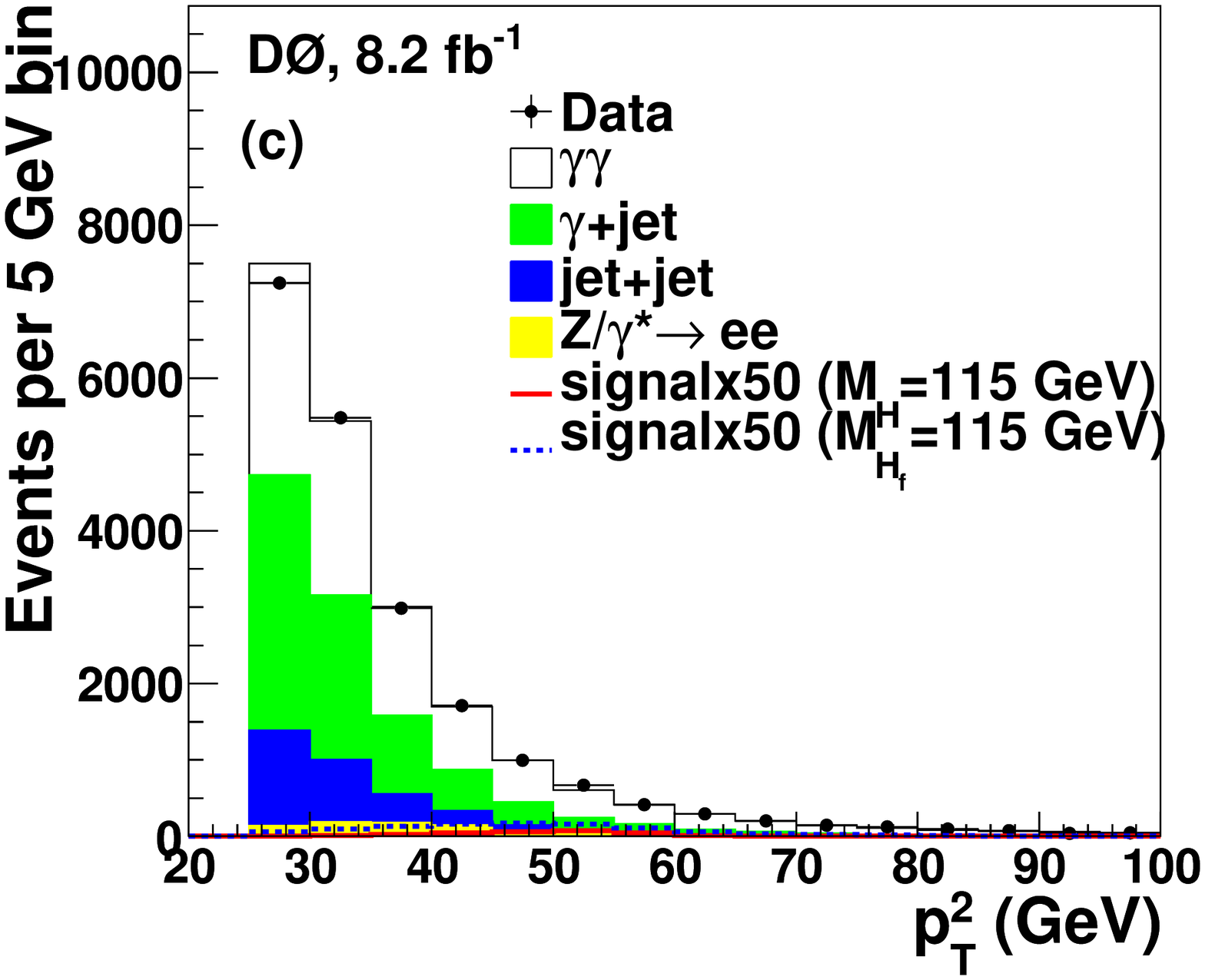}
 \includegraphics[scale=0.4]{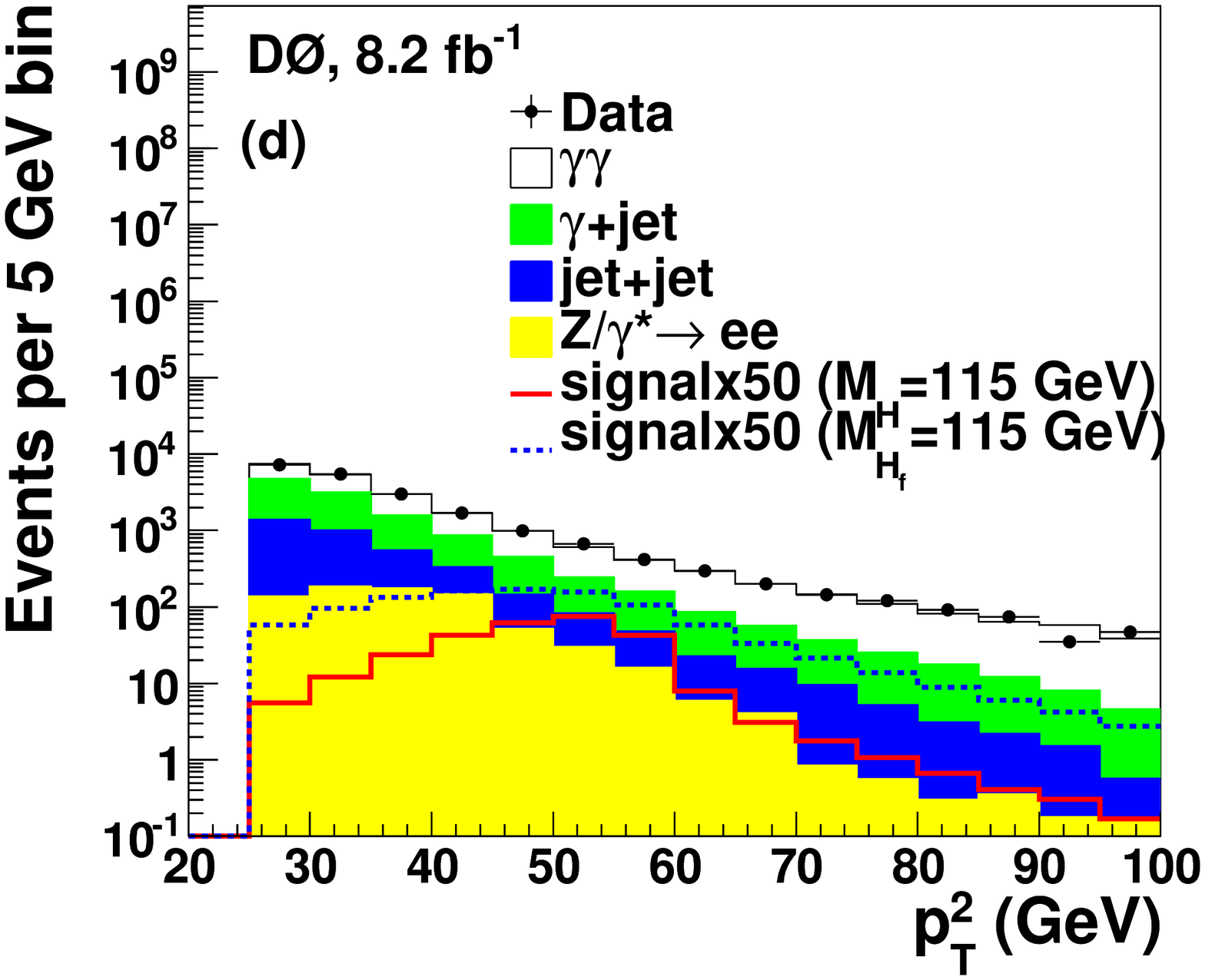} 
   \caption{\small Variables used as input to the BDT: (a, b) leading photon $p_T$ and (c, d) trailing photon $p_T$,
   in linear scale (left column) and logarithmic scale (right column).
   Data are compared to the background prediction. Also shown
   is the expected signal for a SM Higgs boson ($M_H=115$~GeV) and a fermiophobic Higgs boson ($M_{H_{f}}=115$~GeV) both multiplied by a factor of 50. For the fermiophobic Higgs boson, the GF production is absent and the diphoton system is on average more boosted.}   
  \label{plots2}
\end{figure*}

\begin{figure*}[t]
 \centering
\includegraphics[scale=0.4]{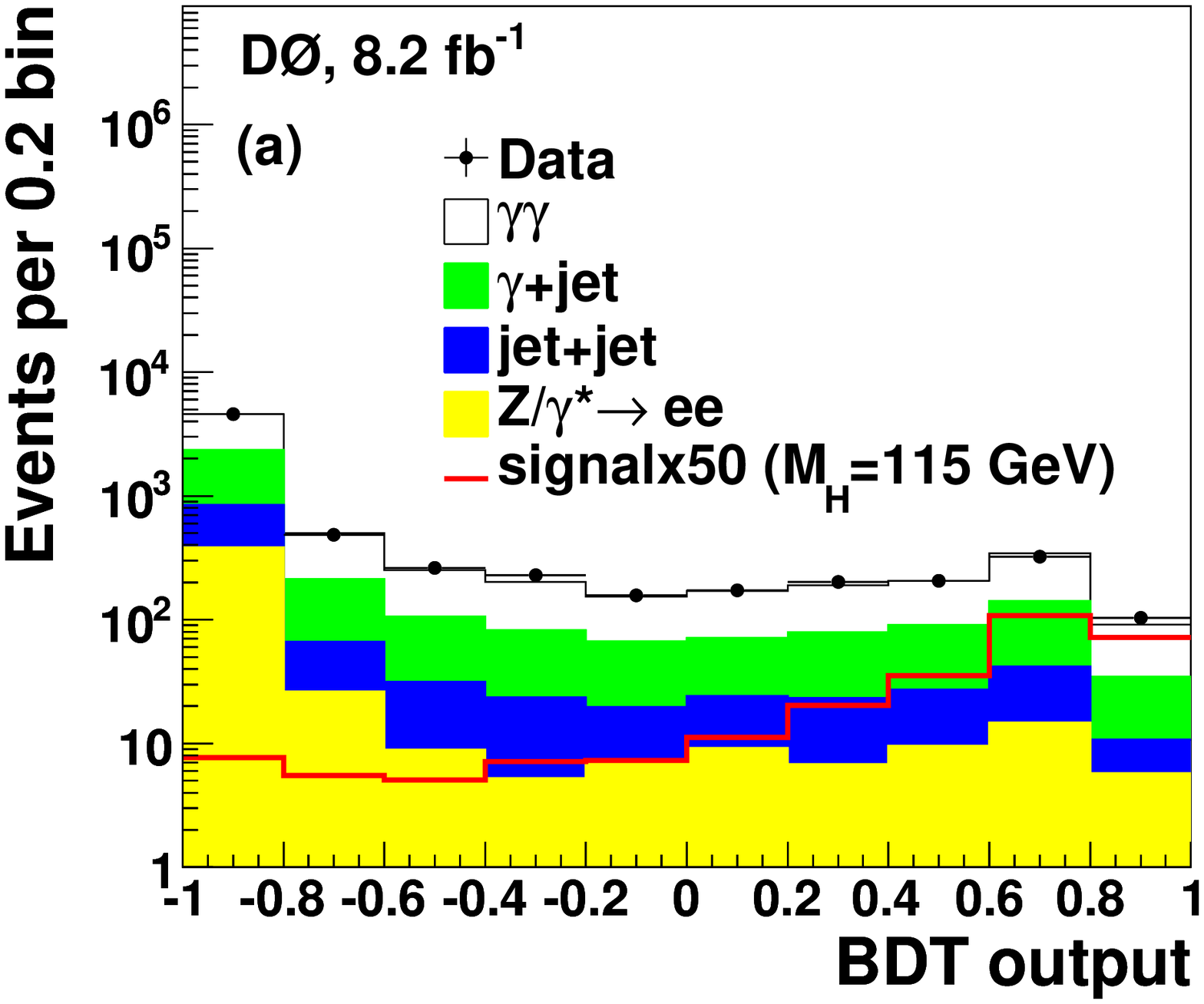}
\includegraphics[scale=0.4]{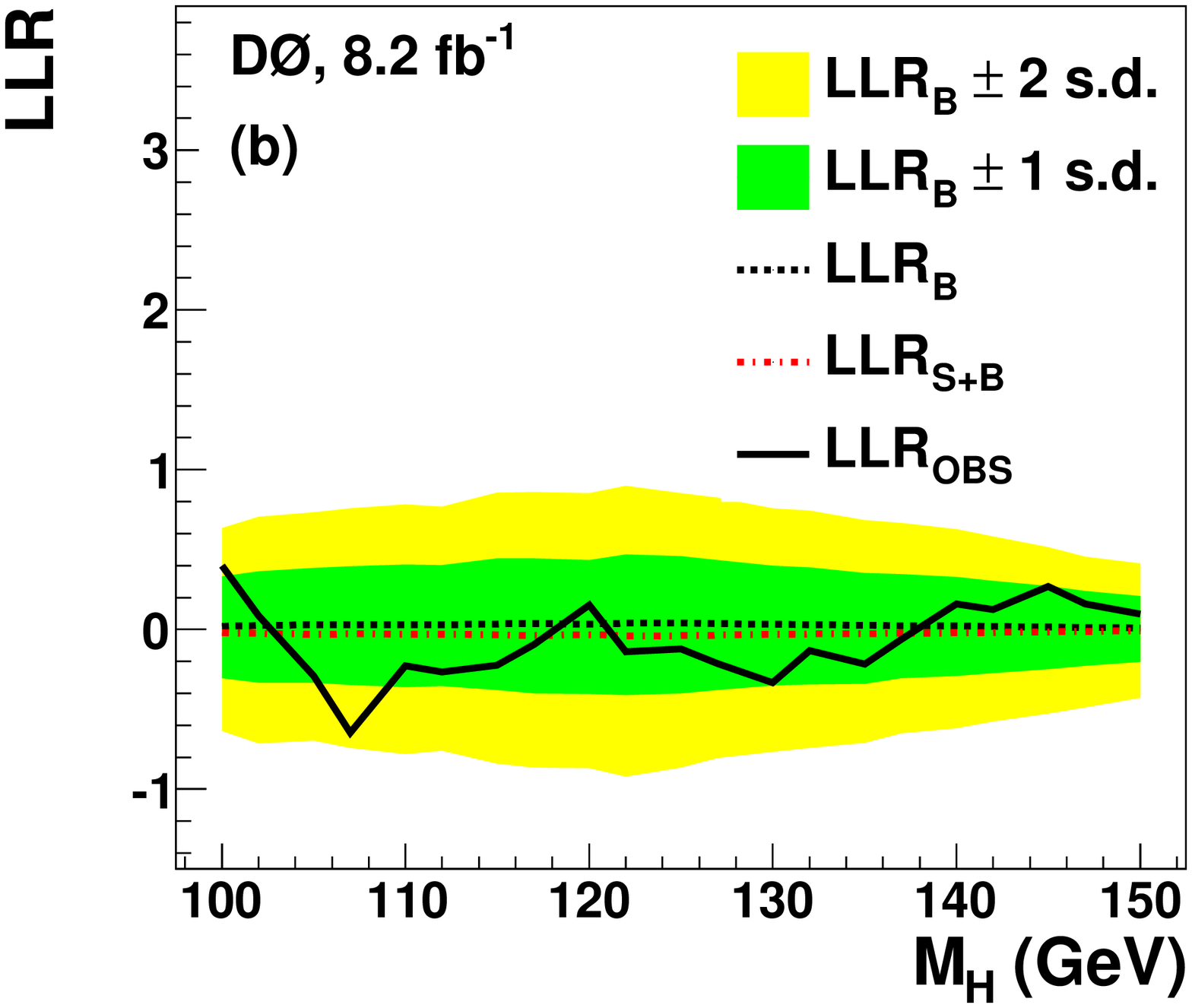}
\includegraphics[scale=0.4]{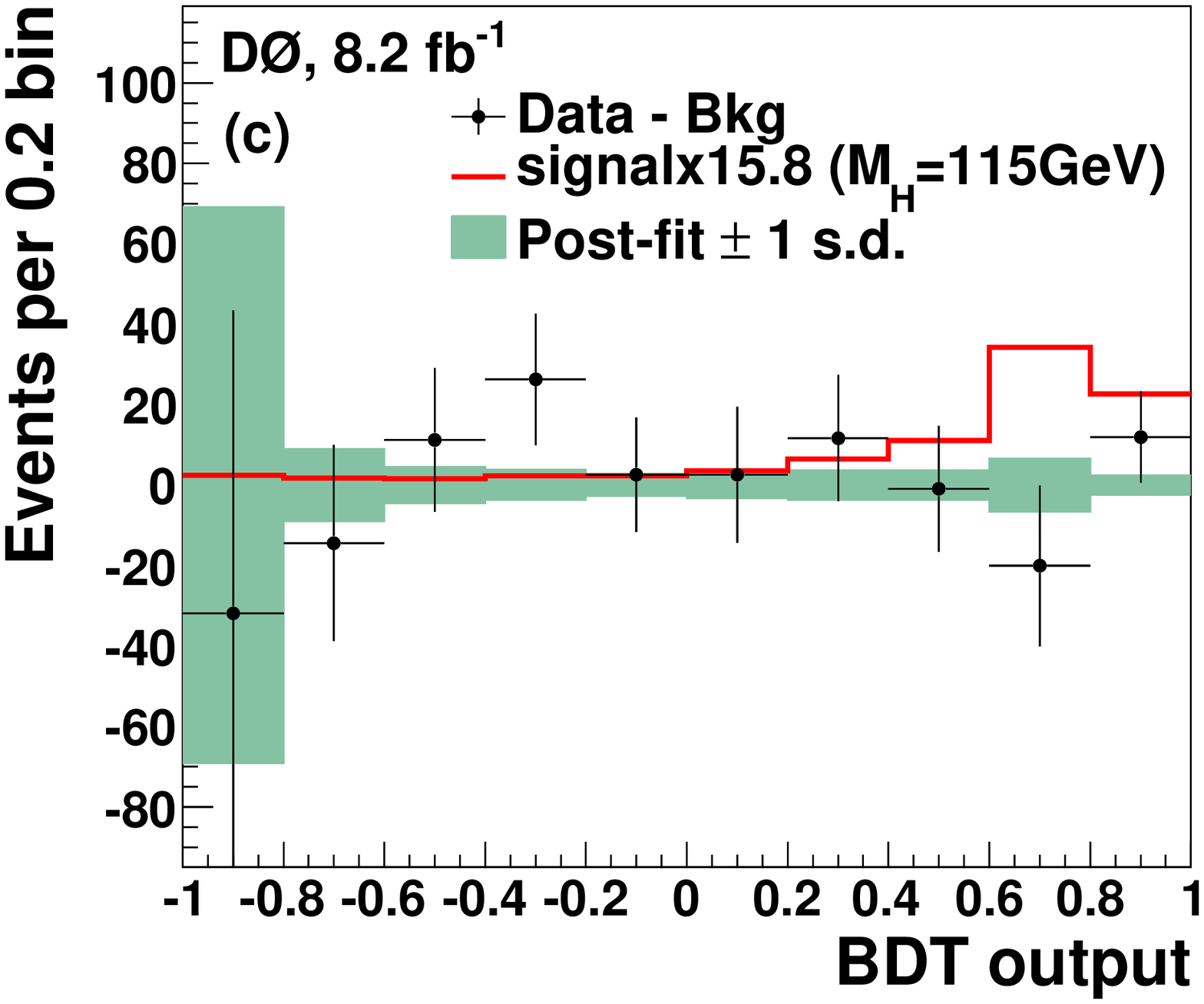}
   \caption{\small SM Higgs search: (a) Distributions of the BDT output after the final selection comparing data to the
background prediction. The expected signal 
multiplied by a factor of 50 is also shown for $M_H=115$~GeV. 
(b) Observed log-likelihood ratio (LLR) as a function of $M_H$
compared to the expected LLR for the background-only hypothesis and signal+background hypothesis. 
The bands correspond to the $\pm$ 1 and $\pm$ 2 standard deviations (s.d.) around the expected LLR
for the background-only hypothesis. 
(c) BDT dependence of the difference between data and expected background for $M_H=115$~GeV. 
    The expected signal is normalized to the observed limit on $\sigma \times B$. The bands represent the post-fit systematic uncertainties.
    }
   \label{plots3}
\end{figure*}

\begin{figure*}[t]
 \centering
\includegraphics[scale=0.4]{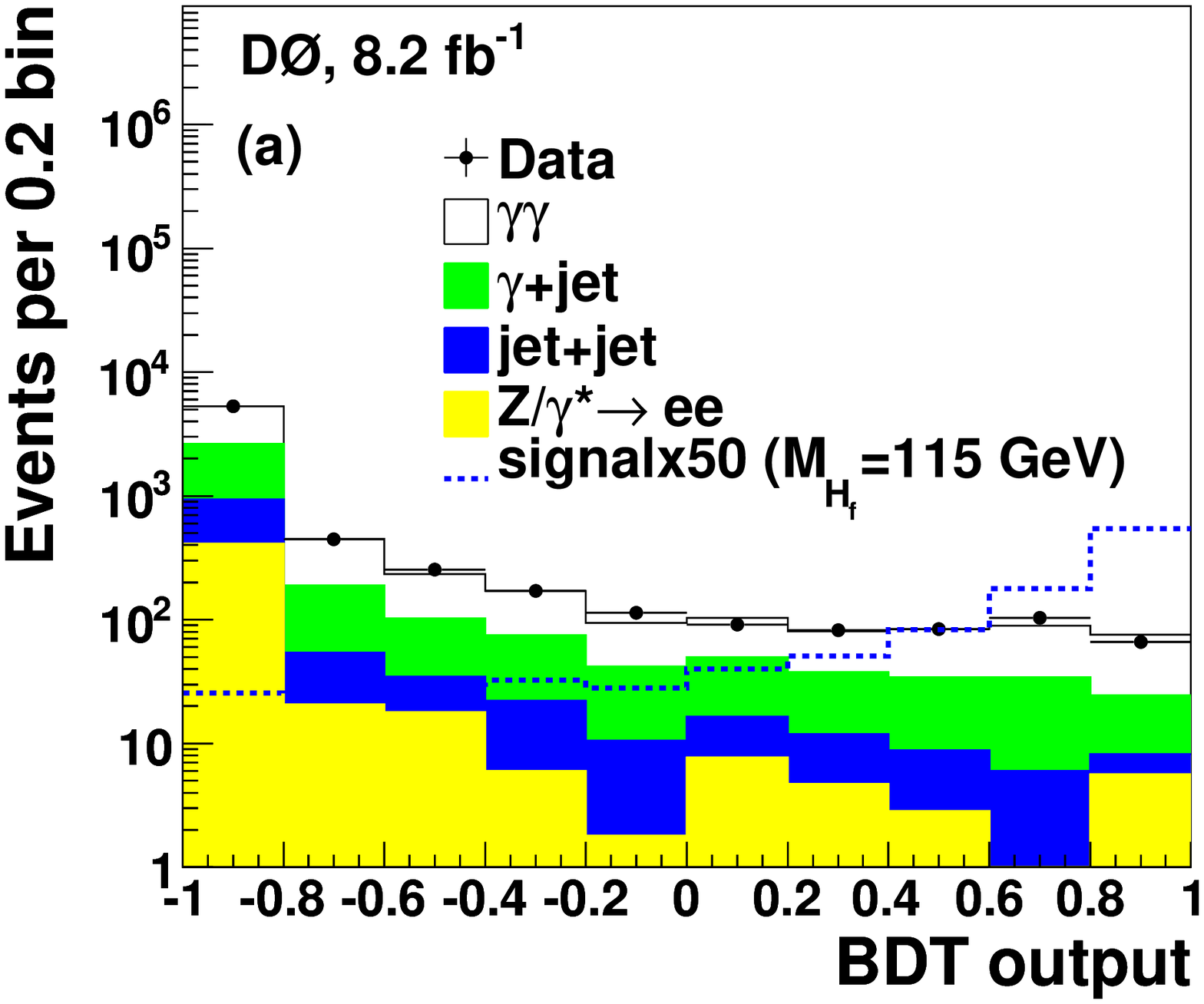}
\includegraphics[scale=0.4]{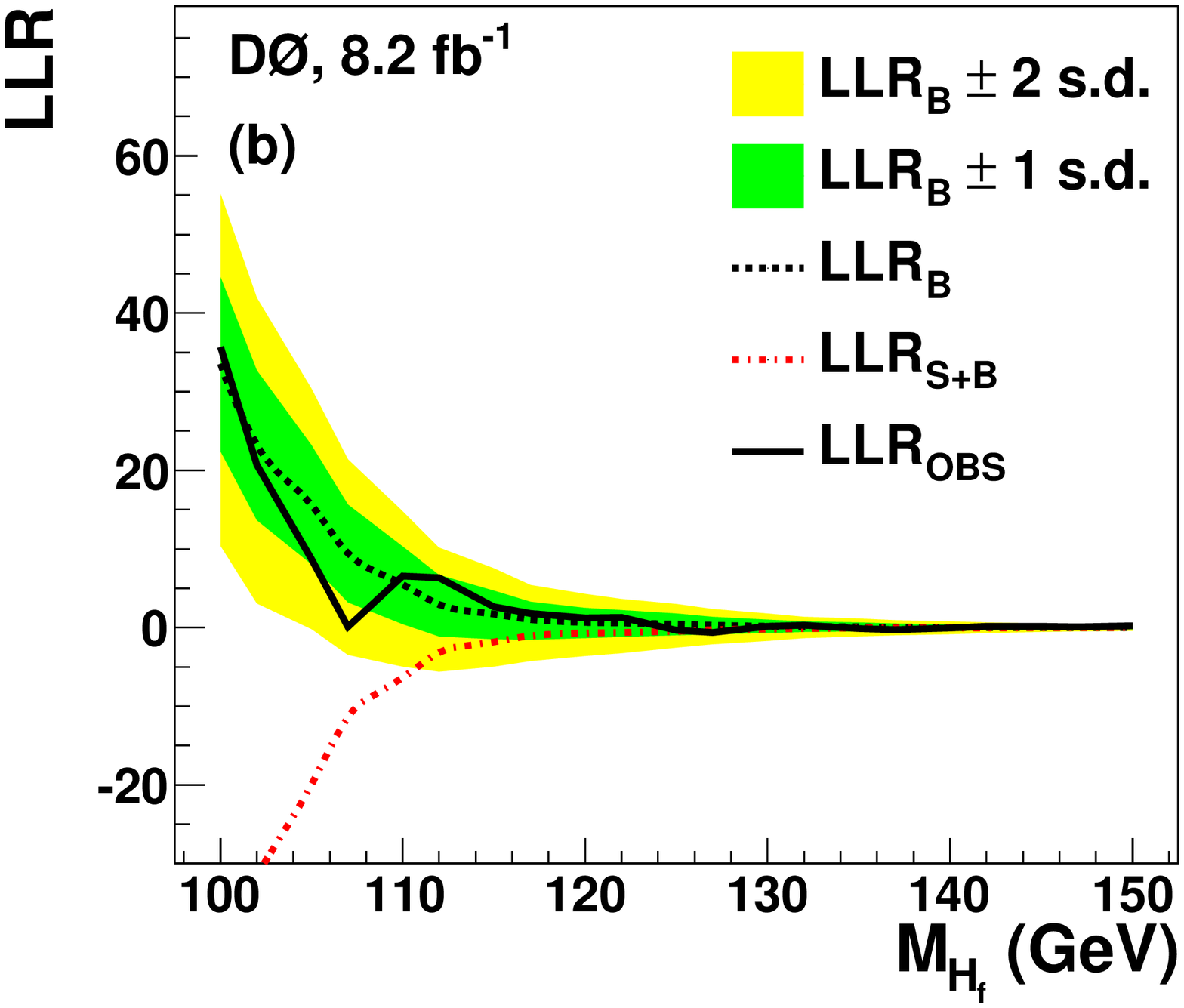}
\includegraphics[scale=0.4]{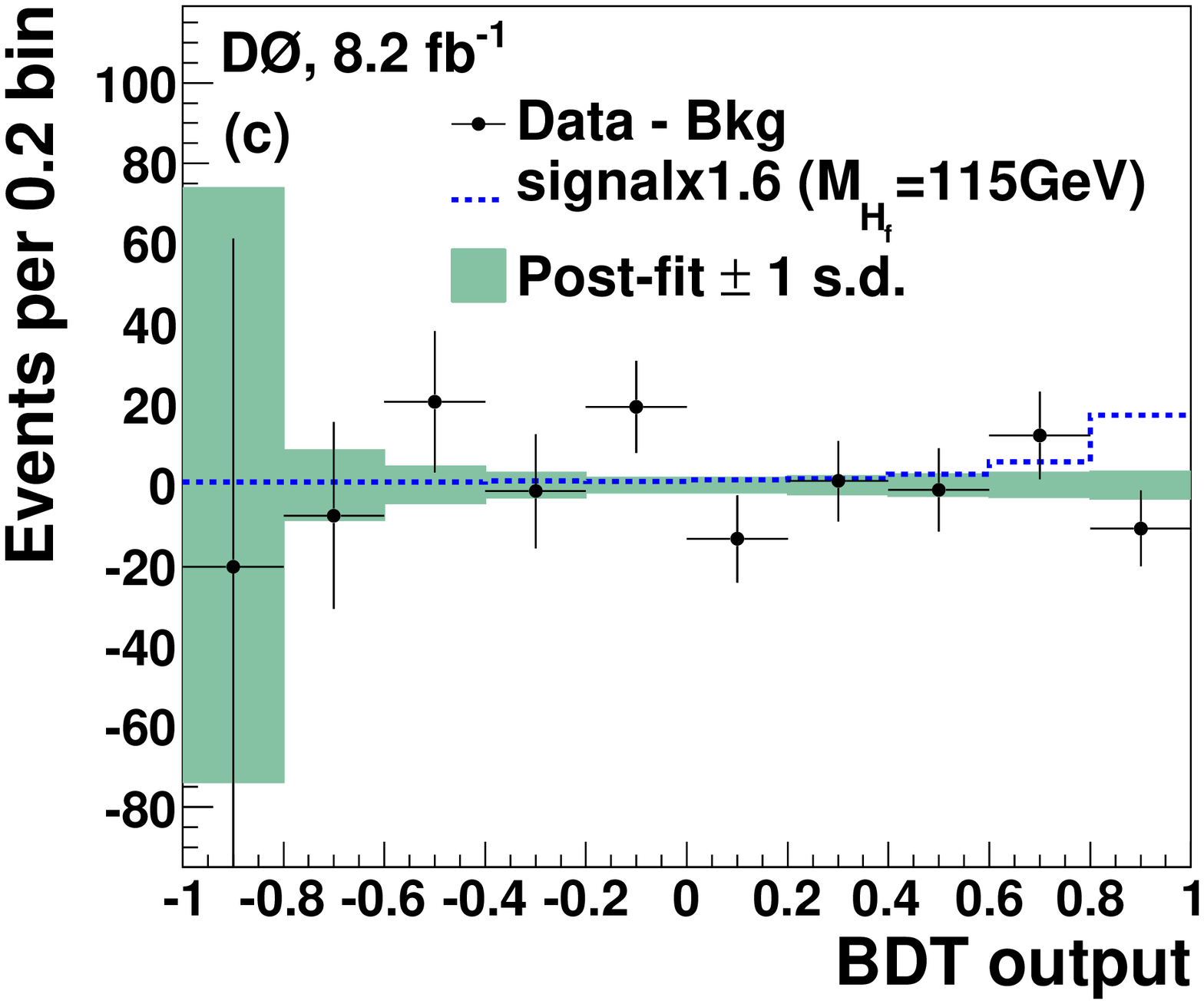}
   \caption{\small Fermiophobic Higgs search: (a) Distributions of the BDT output after final selection comparing data to the
background prediction. The expected signal multiplied by a factor of 50 is shown for $M_{H_{f}}=115$~GeV.
(b) Observed log-likelihood ratio (LLR) as a function of $M_{H_{f}}$
compared to the expected LLRs for the background-only hypothesis and signal+background hypothesis. 
The bands correspond to the $\pm$ 1 and $\pm$ 2 standard deviations (s.d.) around the expected LLR
for the background-only hypothesis.
(c) BDT dependence of the difference between data and expected background for $M_{H_{f}}=115$~GeV.
    The expected signal is normalized to the observed limit on $\sigma \times B$. 
    The bands represent the post-fit systematic uncertainties.
}
  \label{plots4}
\end{figure*}

\begin{table*}
\begin{center}
\caption{
Cross sections for the different Higgs boson production mechanisms and branching fractions
for the SM Higgs boson decays into two photons (${\cal B} (H\to\gamma\gamma)$)
and for fermiophobic Higgs boson decays into two photons (${\cal B} (H_{f}\to\gamma\gamma)$) as a function
of $M_{H}$.}
\vspace{0.2cm}
\label{tab:higgsxsec}
\begin{ruledtabular}
\begin{tabular}{ccccccc}
$M_H$~(GeV) & $\sigma_{gg\rightarrow H}$~(fb) & $\sigma_{WH}$~(fb)  & $\sigma_{ZH}$~(fb)  & $\sigma_{VBF}$~(fb)  & $B(H\rightarrow \gamma\gamma)$~(\%) & $B(H_f \rightarrow \gamma\gamma)$~(\%)\\ 
\hline
   100 &  1821.8   &  291.9    &  168.9      &  100.1 &    0.159 & 18.46     \\
   105 &  1584.7   &  248.4    &  145.9      &    92.3 &    0.178 & 10.42     \\
   110 &  1385.0   &  212.0    &  125.7      &    85.1 &    0.197 & 6.027     \\
   115 &  1215.9   &  174.5    &  103.9      &    78.6 &    0.213 & 3.658     \\
   120 &  1072.3   &  150.1    &    90.2      &    72.7 &     0.225 & 2.334    \\
   125 &    949.3   &  129.5    &    78.5      &     67.1 &    0.230 &1.556      \\
   130 &    842.9   &  112.0    &    68.5      &     62.1 &    0.226 & 1.073      \\
   135 &    750.8   &    97.2    &    60.0      &     57.5 &    0.214 & 0.7586      \\
   140 &    670.6   &    84.6    &    52.7      &     53.2 &    0.194 & 0.5441      \\
   145 &    600.6   &    73.7    &    46.3      &     49.4 &    0.168 & 0.3902      \\
   150 &    539.1   &    64.4    &    40.8      &     45.8 &    0.137 & 0.2733      \\
\end{tabular}
\end{ruledtabular}
\end{center}
\end{table*}

\begin{table*}
\caption{\label{limits_sm_tab} Expected and observed upper limits at 95\% C.L. 
on $\sigma \times {\cal B}$ ($H\to\gamma\gamma$) and on the ratio 
relative to the SM prediction for a SM Higgs boson as a function of $M_{H}$.}
\centering
\begin{tabular}{lccccccccccc}
 \hline \hline
$M_H$ (GeV) & 100  &102.5 &105 &107.5& 110 &112.5& 115&117.5 & 120&122.5 &125 \\
\hline
Expected $\sigma \times B$ (fb) & 61.9 & 55.3 & 51.4 & 48.7 & 47.9 & 46.6 & 42.0 & 38.9 & 40.1 & 36.7 & 33.7   \\
Observed $\sigma \times B$ (fb) & 41.6 & 52.0 & 73.2 & 91.0 & 61.8 & 62.7 & 52.9 & 44.4 & 36.8 & 44.3 & 39.1   \\  
\hline
Expected $\sigma \times B$/SM & 16.3 & 14.7 & 13.9 & 13.4 & 13.5 & 13.5 & 12.5 & 12.0 & 12.9 & 12.4 & 12.0   \\
Observed $\sigma \times B$/SM & 11.0 & 13.9 & 19.9 & 25.0 & 17.4 & 18.2 & 15.8 & 13.7 & 11.8 & 15.0 & 13.9   \\  
\hline \hline
$M_H$ (GeV) & 127.5 & 130&132.5 & 135&137.5 & 140&142.5 & 145&147.5 & 150\\
\hline
Expected $\sigma \times B$ (fb) & 31.7 & 31.0 & 30.4 & 28.4 & 27.2 & 25.7 & 24.5 & 23.8 & 22.5 & 21.6 \\
Observed $\sigma \times B$ (fb) & 40.3 & 42.5 & 35.4 & 36.0 & 29.3 & 21.4 & 21.8 & 16.9 & 18.8 & 19.3\\  
\hline
Expected $\sigma \times B$/SM  & 12.0 & 12.6 & 13.4 & 13.7 & 14.6 & 15.4 & 16.6 & 18.4 & 20.2 & 22.9\\
Observed $\sigma \times B$/SM  & 15.3 & 17.3 & 15.7 & 17.4 & 15.7 & 12.8 & 14.7 & 13.1 & 16.8 & 20.4\\  
\hline\hline 
\end{tabular}
\end{table*}

\begin{table*}
\caption{\label{limits_fh_tab}  Expected and observed upper limits at 95\% C.L. on the cross section times branching ratio for $H_f\to\gamma\gamma$ ($\sigma \times {\cal B}$) and
on the branching ratio ($\cal B$) for a fermiophobic Higgs boson as a function of Higgs boson mass.}
\centering
\begin{tabular}{lccccccccccc}
 \hline \hline
$M_{H_{f}}$ (GeV) & 100  &102.5 &105 &107.5& 110 &112.5& 115&117.5 & 120&122.5 &125 \\
\hline
Expected $\sigma \times B$ (fb) &29.5 &26.3 &23.1 &22.4 &22.5 &23.4 &22.5 &22.8 &20.8 &17.8 &14.8  \\
Observed $\sigma \times B$ (fb) &26.0 &28.4 &31.4 &37.8 &20.7 &16.8 &20.4 &20.2 &18.8 &16.3 &18.7  \\  
\hline
Expected $B$ (\%) &5.2 &5.0 &4.7 &4.9 &5.3 &6.0 &6.3 &6.8 &6.6 &6.1 &5.4    \\
Observed  $B$ (\%)&4.6 &5.4 &6.5 &8.3 &4.9 &4.3 &5.7 &6.0 &6.0 &5.6 &6.8     \\
\hline \hline
$M_{H_{f}}$ (GeV) & 127.5 & 130&132.5 & 135&137.5 & 140&142.5 & 145&147.5 & 150\\
\hline
Expected $\sigma \times B$ (fb)  &14.5 &14.5 &14.5 &13.9 &13.3 &12.7 &12.5 &12.3 &11.4 &11.2 \\
Observed $\sigma \times B$ (fb)  &19.7 &14.9 &13.1 &15.9 &16.2 &13.4 &10.7 &10.1 &9.6 &6.8\\  
\hline
Expected  $B$ (\%) &5.6 &6.0 &6.3 &6.5 &6.6 &6.6 &7.0 &7.3 &7.1 &7.4  \\
Observed  $B$ (\%) &7.6 &6.2 &5.7 &7.4 &8.0 &7.0 &6.0 &6.0 &6.0 &4.5\\
\hline\hline 
\end{tabular}
\end{table*}


\begin{thebibliography}{99}
\bibitem{EtaCov}
Pseudorapidity is defined as $\eta = -\ln[\tan(\theta/2)]$, where 
$\theta$ is the polar angle relative to the proton beam direction and
$\phi$ is the azimuthal angle in the plane transverse to the proton
beam direction.
\end{thebibliography}
\end{document}